\documentclass[suppldata]{interact}

\usepackage{epstopdf}
\usepackage[caption=false]{subfig}
\usepackage[T1]{fontenc}
\usepackage[utf8]{inputenc}
\usepackage{braket}
\usepackage{csquotes}
\usepackage{float}
\usepackage{multirow}
\usepackage{dcolumn}
\usepackage{hyperref}
\usepackage{hhline}
\usepackage{graphicx}
\expandafter\let\csname equation*\endcsname\relax
\expandafter\let\csname endequation*\endcsname\relax
\usepackage{amsmath}
\usepackage{newtxtext}
\DeclareFontEncoding{LS2}{}{}
\DeclareFontSubstitution{LS2}{stix}{m}{n}
\DeclareMathAlphabet\stixcal{LS2}{stixcal}{m} {n}
\usepackage{newtxtext}
\usepackage{newtxmath}
\usepackage{mathtools}
\usepackage{newtxtext}
\usepackage{newtxmath}
\usepackage[mathscr]{eucal}
\usepackage{bm}
\usepackage{graphics}
\usepackage{float}

\usepackage[doi=false,sorting=none,isbn=false,url=false, eprint=false,style=numeric-comp]{biblatex}
\renewbibmacro{in:}{}

\theoremstyle{plain}

\theoremstyle{definition}

\theoremstyle{remark}

\begin{document}
\title{Preservation and enhancement of quantum correlations under Stark effect}

\author{ Nitish Kumar Chandra$^{1}$\thanks{ E-mail: \href{mailto:nitishkrchandra@gmail.com}{nitishkrchandra@gmail.com}}, 
Rajiuddin Sk $^{1}$\thanks{Corresponding E-mail: \href{mailto:skrajiuddin@gmail.com}{skrajiuddin@gmail.com}},
Prasanta K. Panigrahi$^{1}$\thanks{E-mail: \href{mailto:panigrahi.iiser@gmail.com} {panigrahi.iiser@gmail.com}} \\
\vspace{2pt}
\affil{\textsuperscript{1}Department of Physical Sciences,\\ Indian Institute of Science Education and Research Kolkata, India}
}
\maketitle

\vspace{10pt}
  
\begin{abstract}
We analyze the dynamics of quantum correlations by obtaining the exact expression of Bures distance entanglement, trace distance discord, and local quantum uncertainty of two two-level atoms. Here, the atoms undergo two-photon transitions mediated through an intermediate virtual state where each atom is separately coupled to a dissipative reservoir at zero temperature in the presence of the Stark shift effect. We have investigated the dynamics of this atomic system for two different initial conditions of the environment. In the first case, we have assumed the environment's state to be in ground state and in the other case, we have assumed the state to be in first excited state. The second initial condition is significant as it shows the role played by both the Stark shift parameters in contrast to only one of the Stark shift parameters for the first initial condition. Our results demonstrate that quantum correlations can be sustained for an extended period in the presence of Stark shift effect in the case of both Markovian and non-Markovian reservoirs. The effect in the non-Markovian reservoir is more prominent than the Markovian reservoir, even for a very small value of the Stark shift parameter. We observe that among the correlation measures considered, only local quantum uncertainty is accompanied by a sudden change phenomenon, i.e., an abrupt change in the decay rate of a correlation measure. Our findings are significant as preserving quantum correlations is one of the essential aspects in attaining optimum performance in quantum information tasks. 
\end{abstract}


\section{Introduction}

Quantum correlations, a distinguishing feature of quantum systems, are a crucial resource for quantum information science \cite{bell1964einstein}. Their presence in any bipartite or multipartite quantum system results in non-local interactions between the subsystems. Quantum entanglement, one of the most widely considered quantum correlations, is a useful resource for several quantum tasks like teleportation \cite{bennett1993teleporting}, quantum key distribution \cite{ekert1991quantum}, quantum metrology \cite{maccone2013intuitive}, and several others \cite{horodecki2009quantum}. The presence of nonclassical correlations, even in certain separable states, has attracted considerable attention to understand their origin and usefulness as quantum resources \cite{adesso2016measures}. For mixed states, entanglement fails to account for all non-local correlations, resulting in a slew of works devoted to introducing quantum correlation quantifiers beyond entanglement \cite{bera2017quantum}. The entropic quantum discord is one such fundamental quantifier that has gained significant interest in the literature due to its role in quantum speedups using deterministic quantum computation with one quantum bits \cite{knill1998power, datta2008quantum} and other quantum tasks \cite{fanchini2017lectures}. It can be experimentally estimated as a function of temperature for an antiferromagnetic Heisenberg system \cite{singh2015experimental}. However, computing entropic quantum discord for a two-qubit system is NP-complete; hence very hard to calculate for a general case. Only partial findings in the form of closed expressions exist for a few two-qubit states \cite{bera2017quantum}. As a result, geometric versions of quantum discord with various norms have been developed. These geometric discord measures offer easier computability and have found operational significance in quantum tasks such as state distinguishability \cite{spehner2014quantum}. For special cases like two-qubit X state, the exact form of trace distance discord is available in closed form \cite{ciccarello2014toward}.

Recently, Girolami et al. \cite{girolami2013characterizing} proposed a computable discord type measure based on local quantum uncertainty (LQU). It uses skew information and fulfills all the conditions to be a physical quantum correlation measure \cite{luo2003wigner}. The calculation of this measure avoids the difficult step of optimization and fulfills contractivity, which is not adequately proven by geometric discord based on the Hilbert-Schmidt norm \cite{piani2012problem}. In particular, LQU plays a significant role in determining the critical points of quantum phase transitions in multipartite spin systems better than quantum discord and exhibits correlations that are not captured by quantum entanglement, or quantum discord \cite{guo2015examining}. LQU is also associated with quantum Fisher information, which is known to play an essential role in quantum parameter estimation \cite{luo2004wigner}.

Understanding the impact of decoherence on the quantum system is an essential component of quantum information research. In general, the dynamics of systems interacting with their environment are immensely complicated to solve. The quantum system is classified as either an open quantum system or a closed quantum system based on the type of system-environment interaction. Memory effects are a crucial aspect of non-Markovian quantum behavior and are implied by information flow from the surrounding environment to the open system. The process is Markovian if the information flow is unidirectional from the open system to the environment. The Markovian process is based on the premise that the system's characteristic times are significantly longer than the environment's and presupposes that there is always a weak coupling between the system and environment. The dynamics of quantum correlations based on Bures norm, Trace norm, entropy, Fisher information, and quantum uncertainty have been investigated under various noise in both Markovian and non-Markovian regimes \cite{mohamed2021quantum,chen2021dynamics,khedif2018local,slaoui2018dynamics,sk2022protecting,chandra2022dissipative}. 

Quantum correlations are very sensitive to environmental disturbances and decay quickly in the presence of noise, resulting in the phenomenon of entanglement sudden death \cite{yu2004finite,eberly2007end,almeida2007environment,yu2009sudden}. However, for a non-Markovian environment, the revival and oscillation of entanglement are observed owing to the memory effect of the environment \cite{bellomo2007non}. Due to the relevance of entanglement in various applications, numerous techniques like quantum Zeno effect \cite{Joos2009}, weak measurement \cite{sun2010reversing,kim2012protecting}, decoherence-free subspace method \cite{Lidar2003} and others \cite{flores2015two, mortezapour2017protecting} have been proposed to prevent the system to decohere.

The atomic two-photon excitation mechanism, governed by time-dependent quantum dynamics, has been explored in the context of spectroscopic precision in Ref. \cite{haas2006two}. Further, the dynamic Stark shift for two-photon transitions in bound two-body Coulomb systems is computed showing high quantum efficiency in resonant ionisation spectroscopy. In Ref. \cite{agarwal2004dc}, the authors show the application of a dc field based on Stark shifts to regulate spontaneous emission in a cavity. The impact of the Stark effect on the entanglement dynamics in atomic systems with a quantized cavity field was investigated in some works without considering the system's dissipation \cite{ghosh2008control, baghshahi2014entanglement}. Recently, in a dissipative environment, the entanglement dynamics of two two-level atoms were investigated and shown that entanglement captured by concurrence \cite{wootters1998entanglement} can be protected in the presence of Stark shift \cite{golkar2018dynamics}. We show that correlations beyond entanglement can be effectively protected in the presence of Stark shift for two-level atoms having two-photon transitions, where each atom is separately attached to a dissipative reservoir at zero temperature. Our investigations indicate that by altering the Stark shift in the two-qubit system, the quantum correlations captured by Bures distance entanglement, trace distance discord, and local quantum uncertainty can be enhanced for both Markovian and non-Markovian reservoirs. The enhancement of quantum correlations is more effective in the case of a non-Markovian reservoir in comparison to a Markovian reservoir for a specific magnitude of the Stark shift parameter.

The paper is organized as follows: In Section $2$, we briefly review the definition of quantum correlation measures. In Section $3$, we show the dynamics of the two-qubit system's correlation measures in Markovian and non-Markovian settings under the influence of the Stark shift effect when the environments are initially in vacuum states. In Section $4$, we discuss the dynamics of correlation measures when the environments are in the first excited states. Finally, in Section $5$, we conclude by summarizing the results of the paper. 

\section{Quantum Correlation Measures}
In the following, we review the definitions of Bures distance entanglement, Trace distance discord, and local quantum uncertainty, which are robust and reliable measures of quantum correlations.
\subsection{\textbf{Bures Distance Entanglement}}

The Bures distance entanglement measure is defined using Bures distance norm, \cite{streltsov2010linking}  is given by
\begin{equation}
B(t)=\sqrt{2-\sqrt{2+2 \sqrt{1-\mathcal{C}(t)^{2}}}},
\end{equation}
where $\mathcal{C}(t)$ is concurrence, defined as \cite{wootters1998entanglement}
\begin{equation}
\mathcal{C}(t)=\max \left\{0, \sqrt{\lambda_{1}}-\sqrt{\lambda_{2}}-\sqrt{\lambda_{3}}-\sqrt{\lambda_{4}}\right\}.
\end{equation}
Here  $\lambda_{1}\geq\lambda_{2}\geq\lambda_{3}\geq\lambda_{4}$ are the eigenvalues of the matrix, $ T=\rho_{AB}\left(\sigma_{y} \otimes \sigma_{y}\right) \rho^{*}_{AB}\left(\sigma_{y} \otimes \sigma_{y}\right)$, for the bi-partite density matrix $\rho_{AB}$. When $B(t)=\sqrt{2-\sqrt{2}}$, the two-qubit state corresponds to a maximally entangled state, and for a separable state, $B(t)=0$.

\subsection{\textbf{Trace Distance Discord}}
Trace distance discord is another measure of quantum correlation which will be used in our work. It quantifies quantum correlations based on Schatten 1-norm of a given state $\rho$, from the zero discord classical quantum state $\rho_{\mathrm{CQ}}$ :
\begin{equation}
\mathscr{D}(\rho)=\min _{\rho_{\mathrm{CQ}} \in \Delta_{0}}\left\|\rho-\rho_{\mathrm{CQ}}\right\|_{1},
\end{equation}
where $\Delta_{0}$ is the set of classical-quantum states having zero discord and $\mathcal|{X}\|_{1}=$ $\operatorname{Tr}\left(\sqrt{\mathcal{X}^{\dagger} \mathcal{X}}\right)$ is the trace norm of an operator $\mathcal{X}$. The zero-discord classical-quantum states $\rho_{{CQ}}$ can be written as
\begin{equation}
\rho_{\mathrm{CQ}}=\sum_{i=1}^{2} p_{i} \Pi_{i}^{P} \otimes \rho_{i}^{Q},
\end{equation}
where $\left\{p_{i}\right\}$ is the probability distributions with $0 \leq p_{i} \leq 1; \left\{\Pi_{i}^{A}\right\}$ represents a set of orthogonal projectors corresponding to the subsystem $P$ and $\rho_{i}^{Q}$ is the reduced density matrix of the subsystem $Q$.
The closed expression of trace distance discord for a two-qubit $X$ state is given by \cite{ciccarello2014toward}
\begin{equation}\label{TDDEQ}
\mathscr{D}\left(\rho_{P Q}\right)=\sqrt{\frac{\mathcal{R}_{11}^{2} \mathcal{R}_{\max }^{2}-\mathcal{R}_{22}^{2} \mathcal{R}_{\mathrm{min}}^{2}}{\mathcal{R}_{\max }^{2}-\mathcal{R}_{\min }^{2}+\mathcal{R}_{11}^{2}-\mathcal{R}_{22}^{2}}}.
\end{equation}
We have written the above expression in the Fano-Bloch representation \cite{fano1957description}, using which one can write the density matrix $\rho_{PQ}^{\prime}$ as
\begin{equation}
\rho_{P Q}^{\prime}=\frac{1}{4} \sum_{\alpha, \beta=0}^{3} \mathcal{R}_{\alpha \beta} \sigma_{\alpha} \otimes \sigma_{\beta},
\end{equation}
where, $\mathcal{R}_{\alpha \beta}=\operatorname{Tr}\left(\rho_{P Q}^{\prime} \sigma_{\alpha} \otimes \sigma_{\beta}\right)$. 
We know that a bonafide correlation measure is invariant under local transformations, so through suitable local unitary transformations \cite{ciccarello2014toward}, the 
non zero components $\mathcal{R}_{\alpha \beta}$ are given by
\begin{equation}
\begin{array}{ll}
\mathcal{R}_{00}=\operatorname{Tr} \rho=1, & \mathcal{R}_{03}=1-2\left(\rho_{22}+\rho_{44}\right), \quad \mathcal{R}_{30}=1-2\left(\rho_{33}+\rho_{44}\right) \\
\mathcal{R}_{11}=2\left(\left|\rho_{23}\right|+\left|\rho_{14}\right|\right), & \mathcal{R}_{22}=2\left(\left|\rho_{23}\right|-\left|\rho_{14}\right|\right), \quad \mathcal{R}_{33}=1-2\left(\rho_{22}+\rho_{33}\right)
\end{array}
\end{equation}
with $\mathcal{R}_{\max }^{2}=\max \left\{\mathcal{R}_{33}^{2}, \mathcal{R}_{22}^{2}+\mathcal{R}_{30}^{2}\right\}$ and $\mathcal{R}_{\min }^{2}=\min \left\{\mathcal{R}_{11}^{2}, \mathcal{R}_{33}^{2}\right\}$.

\subsection{\textbf{Local Quantum Uncertainty}}
The uncertainty principle limits our capacity to estimate with perfect certainty the measurement outcomes of two incompatible observables at the same time. The uncertainty relations are related to the unique characteristics of quantum mechanics like non-locality, and quantum correlations \cite{luo2003wigner,luo2019quantifying}. Local quantum uncertainty is a quantum correlation measure that captures the minimal quantum uncertainty in a quantum state as a result of local measurements on one of the subsystems of a composite quantum system. It is a discord type measure that offers easier computability, with exact expressions available for qubit-qudit systems compared to the discord measures, which involves a difficult step of optimization over all possible projective measurements, and closed-form is available for just a few forms of two-qubit states \cite{girolami2013characterizing}.
The total uncertainty caused by the effect of measuring a single observable $K$, on a quantum state $\rho$ is given by variance,
\begin{equation}
\operatorname{Var}(\rho, H)=\operatorname{Tr}(\rho K^{2})-(\operatorname{Tr}(\rho K))^{2}.
\end{equation}
Both classical and quantum counterparts have their contribution in this uncertainty. Wigner and Yanase developed the idea of skew information to calculate the quantum counterpart of variance \cite{wigner1997information}. The skew information of a bipartite density matrix $\rho$ is given by
\begin{equation}
I(\rho, K)=-\frac{1}{2} \operatorname{Tr}([\sqrt{\rho}, K]^{2}),
\end{equation}
where [.] is the commutator. It is worth noting that, unlike variance, Wigner-Yanase skew information (WYSI) is unaffected by classical mixing. Girolami et al. defined a measure to quantify quantum correlation based on the definition of WYSI \cite{girolami2013characterizing}. It possesses all of the necessary features for quantifying quantum correlations for a bipartite system. This measure is termed as `Local Quantum Uncertainty' and is evaluated by minimizing the WYSI over the local observables of subsystem $A$,
\begin{equation}
\mathcal{Q}(\rho)=\min _{K_{A}} \mathcal{I}\left(\rho, K_{A} \otimes I_{B}\right),
\end{equation}
where $K_{A}$ is an observable operating on subsystem $A$. The expression of $\mathrm{LQU}$ is given by
\begin{equation}
\mathcal{Q}(\rho)=1-\max \left[\lambda_{1}, \lambda_{2}, \lambda_{3}\right],
\end{equation}
where $\lambda_{i}$ 's are eigenvalues of the $3 \times 3$ matrix $M$ having the elements
\begin{equation}
[M]_{i j} \equiv \operatorname{tr}\left\{\sqrt{\rho}\left(\sigma_{i} \otimes I_{B}\right) \sqrt{\rho}\left(\sigma_{j} \otimes I_{B}\right)\right\},
\end{equation}
where $i, j$ varies from $1,2,3$ and $\sigma_{i}$ 's are Pauli matrices.

In case of pure bipartite states, this correlation measure reduces to linear entropy of the reduced density matrix of the composite system. It becomes zero in the case of classically correlated states and is invariant under local unitary operations.

\section{The Physical Model}\label{Sec3}

We consider a model in which two effective two-level atoms (atom $P$ and atom $Q$) are interacting locally with two independent bosonic reservoirs $\Omega_1$ and $\Omega_2$, respectively in the presence of the Stark effect at zero temperature \cite{golkar2018dynamics}. With the presence of the Stark effect, we can consider this as a qubit system interacting with their environment's field through degenerate two-photon transitions of frequency $2\omega_j$ between the ground state $\ket{0}$ and excited state $\ket{1}$. This two photon transition is mediated by an intermediate level $\ket{i}$, having energy in between the ground state and excited state. The parameters $\eta$ and $\xi$ characterize the Stark shifts of two levels of each atom owing to virtual transitions to the intermediate level.
The two-mode environmental field oscillates with frequencies $\omega_{j_1}$ and $\omega_{j_2}$ while interacting with two two-level atoms. In rotating wave approximation (RWA), the effective Hamiltonian of the system is given by
\begin{align}
\hat{H}_{\mathrm{eff}} = & \omega_{0}\left(\hat{S}_{+}^{P} \hat{S}_{-}^{P}+\hat{S}_{+}^{Q} \hat{S}_{-}^{Q}\right)+\sum_{j_{1}} \omega_{j_{1}} \hat{a}_{j_{1}}^{\dagger} \hat{a}_{j_{1}}+\sum_{j_{2}} \omega_{j_{2}} \hat{a}_{j_{2}}^{\dagger} \hat{a}_{j_{2}}
+\sum_{j_{1}} \gamma_{j_{1}}\left(\hat{a}_{j_{1}}^{\dagger^{2}} \hat{S}_{-}^{P}+\hat{a}_{j_{1}}^{2} \hat{S}_{+}^{P}\right)\nonumber \\
&+\sum_{j_{1}} \hat{a}_{j_{1}}^{\dagger} \hat{a}_{j_{1}}\left(\eta_{j_{1}} \hat{S}_{-}^{P} \hat{S}_{+}^{P}+\xi_{j_{1}} \hat{S}_{+}^{P} \hat{S}_{-}^{P}\right)+\sum_{j_{2}} \gamma_{j_{2}}\left(\hat{a}_{j_{2}}^{\dagger^{2}} \hat{S}_{-}^{Q}+\hat{a}_{j_{2}}^{2} \hat{S}_{+}^{Q}\right)\nonumber \\
&+\sum_{j_{2}} \hat{a}_{j_{2}}^{\dagger} \hat{a}_{j_{2}}\left(\eta_{j_{2}} \hat{S}_{-}^{Q} \hat{S}_{+}^{Q}+\xi_{j_{2}} \hat{S}_{+}^{Q} \hat{S}_{-}^{Q}\right).
\end{align}
Here, the operator $\hat{a}^{\dagger}_{j}$ and $\hat{a}_{j}$ are the creation and annihilation operators of the $j^{th}$ mode of environment and $\hat{S}^{P,Q}_{\pm}$ represent the raising and lowering operators for the qubits describing atoms P and Q. The coefficients $\gamma_{j_{1}}$ and $\gamma_{j_{2}}$ are coupling coefficients describing the two-photon strength of the qubit for the environmental modes $j_1$ and $j_2$ respectively, and the parameters $\eta$ and $\xi$ are Stark shift coefficients \cite{puri1988quantum}. In the interaction picture, the above Hamiltonian takes the following form:
\begin{align}\label{Hint}
\hat{H}_{\mathrm{int}}=& \sum_{j_{1}} \gamma_{j_{1}}\left(\hat{a}_{j_{1}}^{\hat{2}} \hat{S}_{-}^{P} \mathrm{e}^{-\mathrm{i}\left(\omega_{0}-2 \omega_{j_{1}}\right) t}+\hat{a}_{j_{1}}^{2} \hat{S}_{+}^{P} e^{i\left(\omega_{0}-2 \omega_{j_{1}}\right) t}\right)
+\sum_{j_{1}} \hat{a}_{j_{1}}^{\dagger} \hat{a}_{j_{1}}\left(\eta_{j_{1}} \hat{S}_{-}^{P} \hat{S}_{+}^{P}+\xi_{j_{1}} \hat{S}_{+}^{P} \hat{S}_{-}^{P}\right) \nonumber\\
&+\sum_{k_{2}} \gamma_{j_{2}}\left(\hat{a}_{j_{2}}^{\dagger^{2}} \hat{S}_{-}^{Q} \mathrm{e}^{-\mathrm{i}\left(\omega_{0}-2 \omega_{j_{2}}\right)t}+\hat{a}_{j_{2}}^{2} \hat{S}_{+}^{Q}\right) e^{i\left(\omega_{0}-2\omega_{j_{2}}\right)t}+\sum_{j_{2}} \hat{a}_{j_{2}}^{\dagger} \hat{a}_{j_{2}}\left(\eta_{j_{2}} \hat{S}_{-}^{Q} \hat{S}_{+}^{Q}+\xi_{j_{2}} \hat{S}_{+}^{Q} \hat{S}_{-}^{Q}\right).
\end{align}
We take the initial state of the system as an entangled state in the form:
\begin{equation}
|\Phi(0)\rangle=\left(x|0\rangle_{P}|1\rangle_{Q}+\sqrt{1-x^{2}}|1\rangle_{P}|0\rangle_{Q}\right) \otimes\left|0_{j_{1}}\right\rangle_{\Omega_{1}} \mid 0_{j_{2}}\rangle_{\Omega_{2}},
\end{equation}
with $x \in[0,1]$, where $\left|0_{j_{1}}\right\rangle_{\Omega_{1}}$  and $\left|0_{j_{2}}\right\rangle_{\Omega_{2}}$ indicate the vacuum state of the  environment of P and Q, and $\ket{0}$ and $\ket{1}$ are the ground state and excited state of the two-level atoms. The time evolution of the total system for time $t>0$ can be written as
\begin{equation}
 \begin{split}\label{PHIt}   
|\Phi(t)\rangle=(b_1(t)\ket{1}_P\ket{0}_Q+b_2(t)\ket{0}_P\ket{1}_Q)\ket{0_{j_1}}_{\Omega_1}\ket{0_{j2}}_{\Omega_2}+\sum_{k_{1}} b_{j_{1}}(t)|0\rangle_{P}|0\rangle_{Q}\left|2_{j_{1}}\right\rangle_{\Omega_{1}}\left|0_{j_{2}}\right\rangle_{\Omega_{2}}\\
+\sum_{j_{2}} b_{j}(t)|0\rangle_{P}|O\rangle_{Q}\left|0_{j_{1}}\right\rangle_{\Omega_{1}}\left|2_{j_2}\right\rangle_{\Omega_{2}},
\end{split}
\end{equation}
where $\left|2_{j}\right\rangle_{\Omega_{j}}$ implies the presence of two photons in the $j^{th}$ mode. So, each of the environment has two states $\left|0_{j}\right\rangle$ and $\left|2_{j}\right\rangle$. We get the following differential equations of probability amplitude from the Schrodinger equation using Eqs. \ref{Hint} and \ref{PHIt} as
\begin{equation}\label{Eq5}
\dot{b}_{j}(t)=-\mathrm{i} \sqrt{2} \sum_{j_{m}} \gamma_{j_{m}} b_{j_{m}}(t) \mathrm{e}^{\mathrm{i}\left(\omega_{0}-\omega_{j}\right) t}
\end{equation}
and 
\begin{equation}\label{Eq6}
\dot{b}_{j_m}(t)=-\mathrm{i} \sqrt{2} \gamma_{j_{m}}^{*} b_{j}(t) \mathrm{e}^{-\mathrm{i}\left(\omega_{0}-\omega_{j_{m}}\right) t}-2 \mathrm{i} \eta_{j_{m}} b_{j_{m}}(t), m=1,2.
\end{equation}
The parameters $\xi_{j_{1}}$ and $\xi_{j_{2}}$ do not affect the dynamics of the quantum correlations because $\hat{a}_{j_{1}}^{\dagger} \hat{a}_{j_{1}}\xi_{j_{1}} \hat{S}_{+}^{P} \hat{S}_{-}^{P}|\Phi(t)\rangle$ and $\hat{a}_{j_{2}}^{\dagger} \hat{a}_{j_{2}}\xi_{j_{2}} \hat{S}_{+}^{Q} \hat{S}_{-}^{Q}|\Phi(t)\rangle$ becomes zero \cite{golkar2018dynamics}. This happens due to the nature of $|\Phi(t)\rangle$ which is a result of assuming the initial state of the environment to be in ground state. We observe that the parameters $\xi_{j_{1}}$ and $\xi_{j_{2}}$ do not play any role on the dynamics and further on the quantum correlations of the model considered in our work. If we change the initial state of the environments, we can observe the role played by both the Stark shift parameters. After integrating Eq. \ref{Eq6}, and using Eq. \ref{Eq5}, we obtain the following differential equation:
\begin{equation}\label{Eq7}
\dot{b}_{j}(t)=-2 \int_{0}^{t} \mathrm{~d} t^{\prime} g\left(t-t^{\prime}\right) b_{j}\left(t^{\prime}\right),
\end{equation}
where the correlation function can expressed as
\begin{equation}\label{Eq8}
g\left(t-t^{\prime}\right)=\int \mathrm{d} \omega_{j} J\left(\omega_{j}\right) \exp \left[\mathrm{i}\left(\omega_{0}-2 \omega_{k}-2 \eta\right)\left(t-t^{\prime}\right)\right].
\end{equation}
Here we consider $\eta_{j_{1}}=\eta_{j_{2}}=\eta$ and $\omega_{j_{1}}=\omega_{j_{2}}=\omega_{j}$. The discrete sum over the reservoir modes can be approximated using the integral form when there are a large number of reservoir modes, $\sum_{k}\left|\gamma_{j}\right|^{2} \rightarrow \int \mathrm{d} \omega_{j} J\left(\omega_{j}\right)$, where $J\left(\omega_{j}\right)$ represents the electromagnetic field's spectral density within a lossy cavity.

The reservoir's spectral density is assumed to be Lorentzian \cite{breuer2002theory},

\begin{equation}\label{Eq9}
J\left(\omega_{j}\right)=\frac{1}{2 \pi} \frac{\gamma_{0} \Lambda^{2}}{\left(\omega_{0}-2 \omega_{j}\right)^{2}+\Lambda^{2}},
\end{equation}
where the parameter $\Lambda$ indicates the width of Lorentzian distribution and the parameter $\omega_{0}$ is the transition frequency of the atom. The relationship between reservoir correlation time $\tau_{\Omega}$ and $\Lambda$ is $\tau_{\Omega} \approx \Lambda^{-1}$. The decay rate of the excited atom is represented by $\gamma_{0}$, and is directly related with the relaxation time $\tau_{s} \approx \gamma_{0}^{-1}$ \cite{spohn1980kinetic}. The correlation function $g\left(t-t^{\prime}\right)$ obtained using Eqs. \ref{Eq8} and \ref{Eq9} is given by
\begin{equation}\label{Cor}
g\left(t-t^{\prime}\right)=\frac{\gamma_{0} \Lambda}{2} \exp \left[-(\Lambda+2 \mathrm{i} \eta)\left(t-t^{\prime}\right)\right].
\end{equation}
We can find the exact solution $b_1(t)$ and $b_2(t)$
by making use of Laplace's method on Eq. \ref{Eq7}, and using above correlation function: 
\begin{equation}
b_1(t)= x \phi(t), \quad b_2(t)=\sqrt{1-x^{2}} \phi(t).
\end{equation}
where,
\begin{equation}
\phi(t)=e^{-\frac{(\Lambda+2 i \eta) t}{2}}\left[\cosh \left(\frac{\sigma t}{2}\right)+\frac{\Lambda+2 \mathrm{i} \eta}{\sigma} \sinh \left(\frac{\sigma t}{2}\right)\right]
\end{equation},
and
\begin{equation}
\sigma=\sqrt{-4 \gamma_{0} \Lambda+(\Lambda+2 \mathrm{i} \eta)^{2}}.
\end{equation}
If we write, $\Lambda=n_{1}\gamma_{0}$ and $\eta=n_{2}\gamma_{0}$ where $n_{1}$ and $n_{2}$ are some positive numbers. 
By putting in the values of Stark shift coefficients in terms of $\gamma_{0}$,
\begin{equation}
\sigma=\sqrt{-4n_{1} \gamma_{0}+(n_{1}+2 \mathrm{i} n_{2})^{2}}\gamma_{0} = \sigma_{1}\gamma_{0},
\end{equation}
where,
\begin{equation}
\sigma_{1}=\sqrt{-4n_{1} \gamma_{0}+(n_{1}+2 \mathrm{i} n_{2})^{2}}.
\end{equation}
Thus,
\begin{equation}
\phi(\gamma_{0}t)=e^{-\frac{(n_{1}+2 i n_{2}) \gamma_{0}t}{2}}\left[\cosh \left(\frac{\sigma_{1}\gamma_{0} t}{2}\right)+\frac{n_{1}+2 \mathrm{i} n_{2}}{\sigma_{1}} \sinh \left(\frac{\sigma_{1}\gamma_{0} t}{2}\right)\right].
\end{equation}

We note that the scaled time is given by '$\tau = \gamma_{0}t$',
The behavior of $b_1(t)$ and $b_2(t)$ is differentiated by two regimes depending on the weak or strong coupling between the open system and environment \cite{dalton2001theory}. In the weak coupling regime, we have, $\gamma_0<\Lambda/2$ and $\tau_s<2\tau_{\Omega}$. The relaxation time in this region is greater than the reservoir's correlation time, and we observe a decaying process as time progresses. This behavior of the qubit reservoir system is known as Markovian. On the strong coupling regime, we have  $\gamma_0>\Lambda/2$ and $\tau_s>2\tau_{\Omega}$ implying the relaxation time is less than the qubit-reservoir correlation time. The dynamics in this region are known as non-Markovian dynamics, in which memory effect comes into play, and the revival of quantum correlation is observed with damped oscillation. In this work, we investigate both the Markovian and non-Markovian dynamics by considering the Stark shift effect.

\subsection{\textbf{Dynamics of Bures Distance Entanglement}}

The reduced density matrix of the system (consisting of atoms P and Q) in the atomic basis $\left\{|1\rangle_{P}|1\rangle_{Q},|1\rangle_{P}|0\rangle_{Q},|0\rangle_{P}|1\rangle_{Q},|0\rangle_{P}|0\rangle_{Q}\right\}$ is given by
\begin{equation}
    \rho(t)=\left(\begin{array}{cccc}
0 & 0 & 0 & 0 \\
0 & \left|b_{1}(t)\right|^{2} & b_{1}(t) b_{2}^{*}(t) & 0 \\
0 & b_{1}^{*}(t) b_{2}(t) & \left|b_{2}(t)\right|^{2} & 0 \\
0 & 0 & 0 & 1-\left|b_{1}(t)\right|^{2}-\left|b_{2}(t)\right|^{2}.
\end{array}\right)
\end{equation}

Using the procedure of finding concurrence as discussed earlier, for the above  density matrix, one finds
$$
C(t)=2\left|b_{1}(t) b_{2}^{*}(t)\right|,
$$

where the Bures distance entanglement measure is given by
$$
B(t)=\sqrt{2-\sqrt{2+2 \sqrt{1-C(t)^{2}}}}
$$
\begin{figure}[hbt!]
     \centering
\subfloat[]{\includegraphics[width=0.5\textwidth]{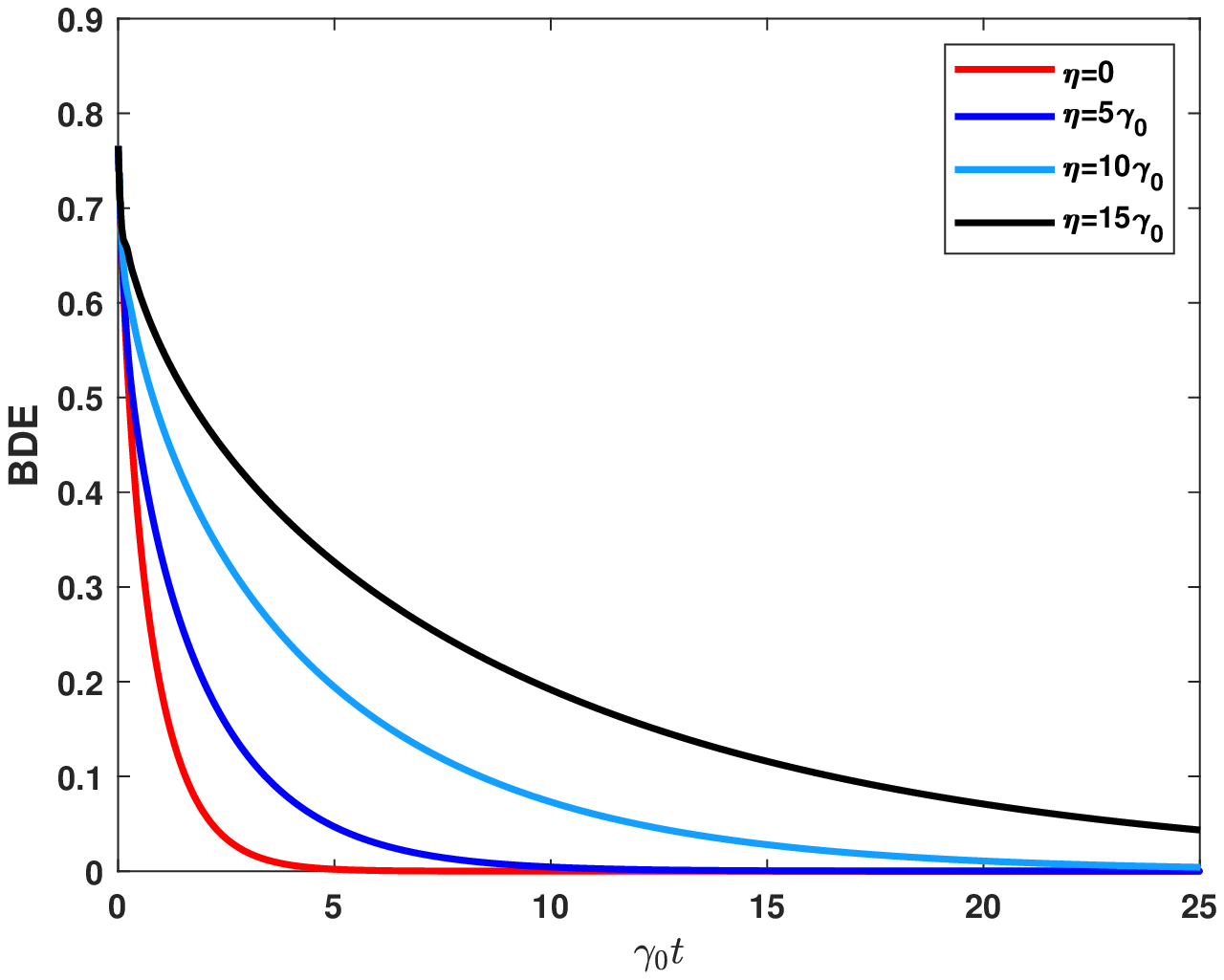}\label{fig1a}}
\subfloat[]{\includegraphics[width=0.5\textwidth]{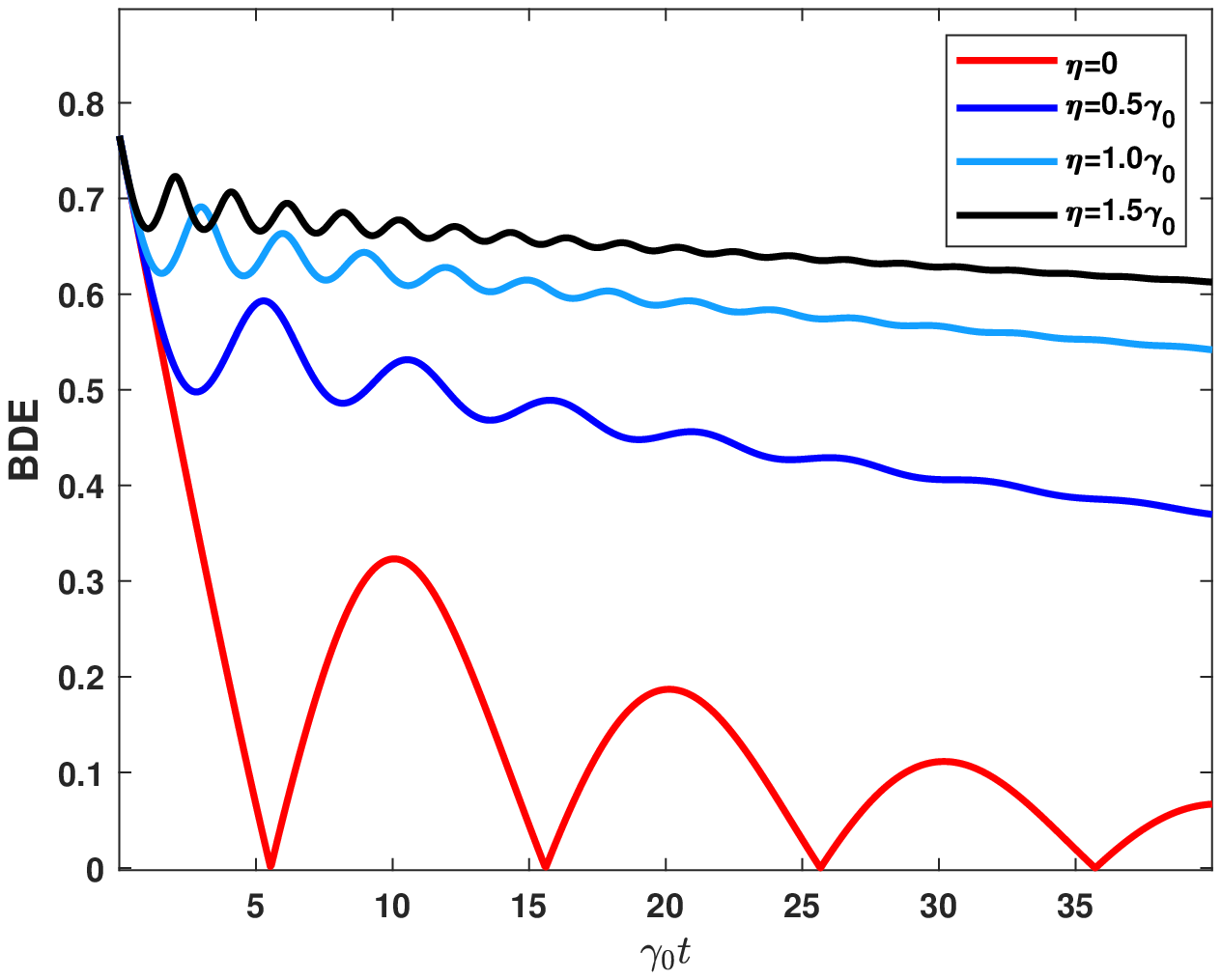}\label{fig1b}}
\caption{Dynamics of Bures distance entanglement in (a) Markovian reservoir for $\Lambda=10\gamma_0$ with Stark shift values $\eta=0,5\gamma_0,10\gamma_0,15\gamma_0$ (b) non-Markovian reservoir for $\Lambda=0.1\gamma_0$, with $\eta=0,0.5\gamma_0,1.0\gamma_0,1.5\gamma_0$.}
\label{fig1}
\end{figure}
\begin{figure}[hbt!]
     \centering
\subfloat[]{\includegraphics[width=0.5\textwidth]{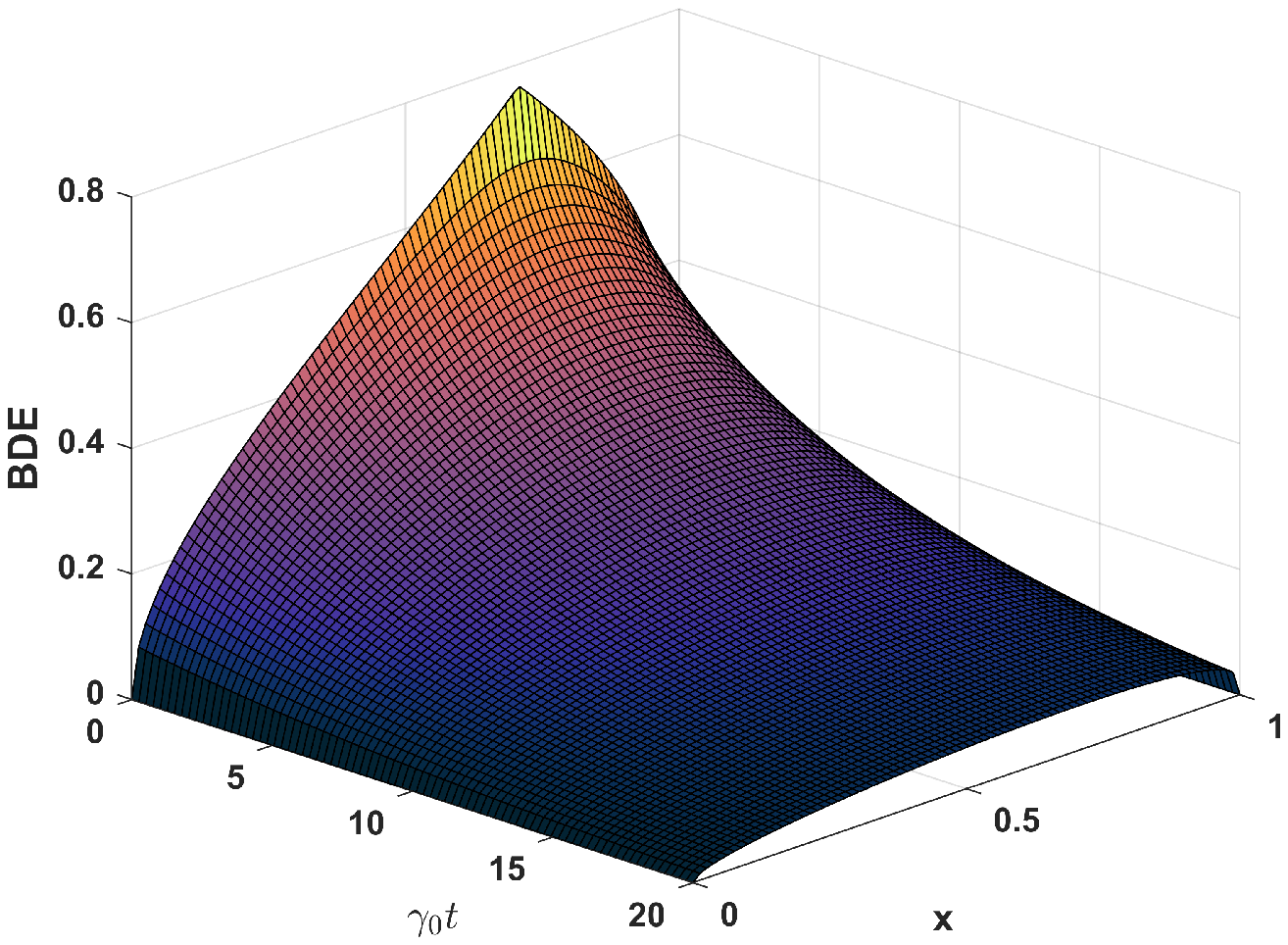}\label{fig2a}}
\subfloat[]{\includegraphics[width=0.5\textwidth]{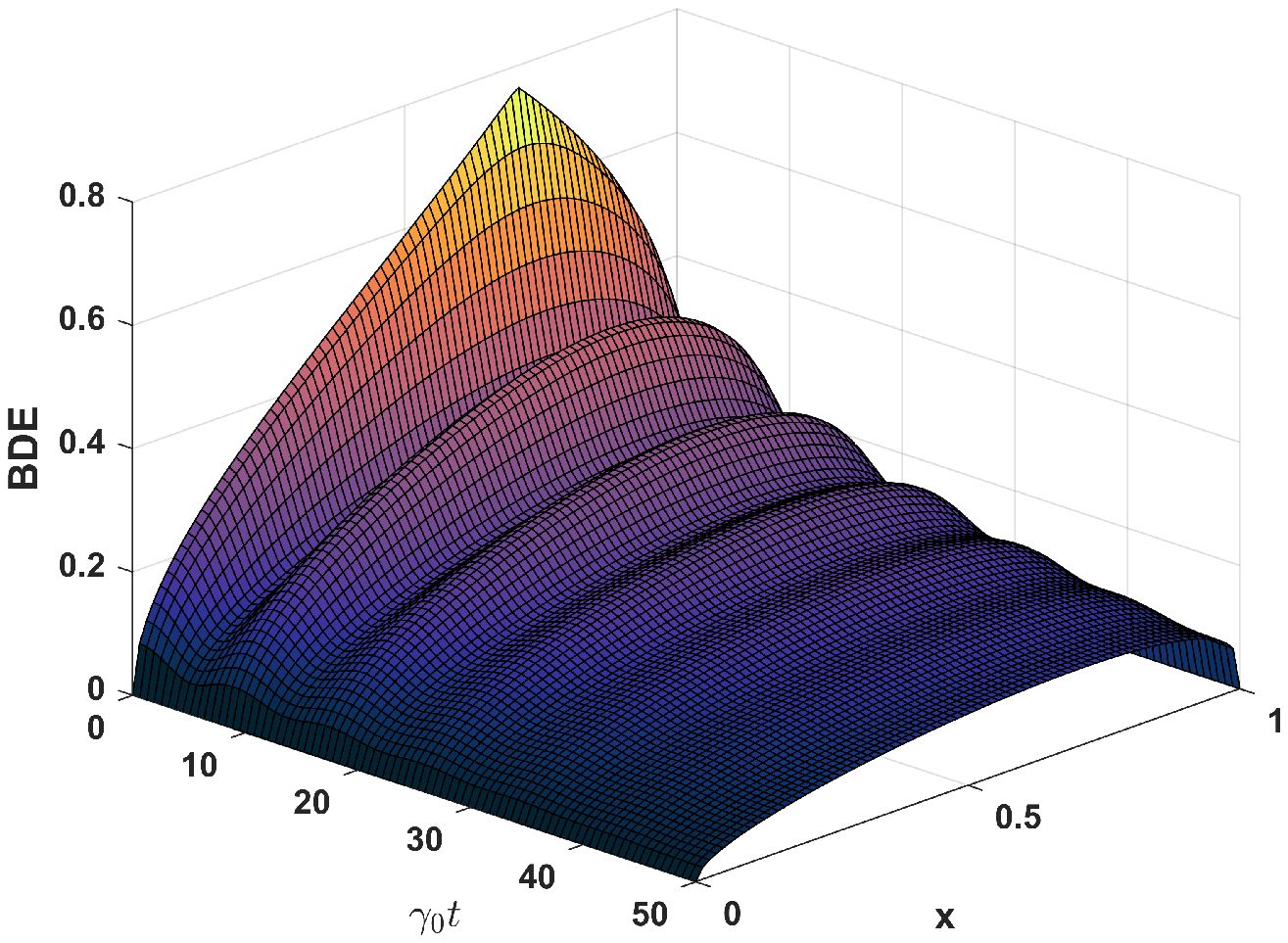}\label{fig2b}}
\caption{The surface plot of BDE with scaled time $\gamma_{0} t$ and state parameter $x$ in (a) in Markovian reservoir for $\Lambda=10\gamma_0$ with the value of Stark shift ($\eta=15\gamma_0$) and (b) in non-Markovian reservoir for $\Lambda=0.1\gamma_0$, with the value of Stark shift parameter ($\eta=0.2\gamma_0$).}
\label{fig2}
\end{figure}
In Fig.\ref{fig1}, we show the temporal dynamics of the Bures distance entanglement (BDE) in the presence and absence of Stark shift for both Markovian and non-Markovian reservoirs. In the Markovian regime, we observe the exponential decay of BDE. However, in the non-Markovian regime, we observe periodic death and revival of BDE. For a particular value of $\Lambda$, BDE is protected more efficiently with the increase of Stark shift parameter $\eta$. Specifically, in non-Markovian regime, the effect of $\eta$ is more significant. The surface plot of BDE with scaled time and state parameter x is depicted in Fig.\ref{fig2}. We observe that for maximally entangled input state ($x=\frac{1}{\sqrt{2}})$, BDE has maximum value of $\sqrt{2-\sqrt{2}}$.
\subsection{\textbf{Dynamics of Trace Distance Discord}}
We can compute trace distance discord for the X state using the expression \ref{TDDEQ}.
\begin{figure}[hbt!]
     \centering
\subfloat[]{\includegraphics[width=0.5\textwidth]{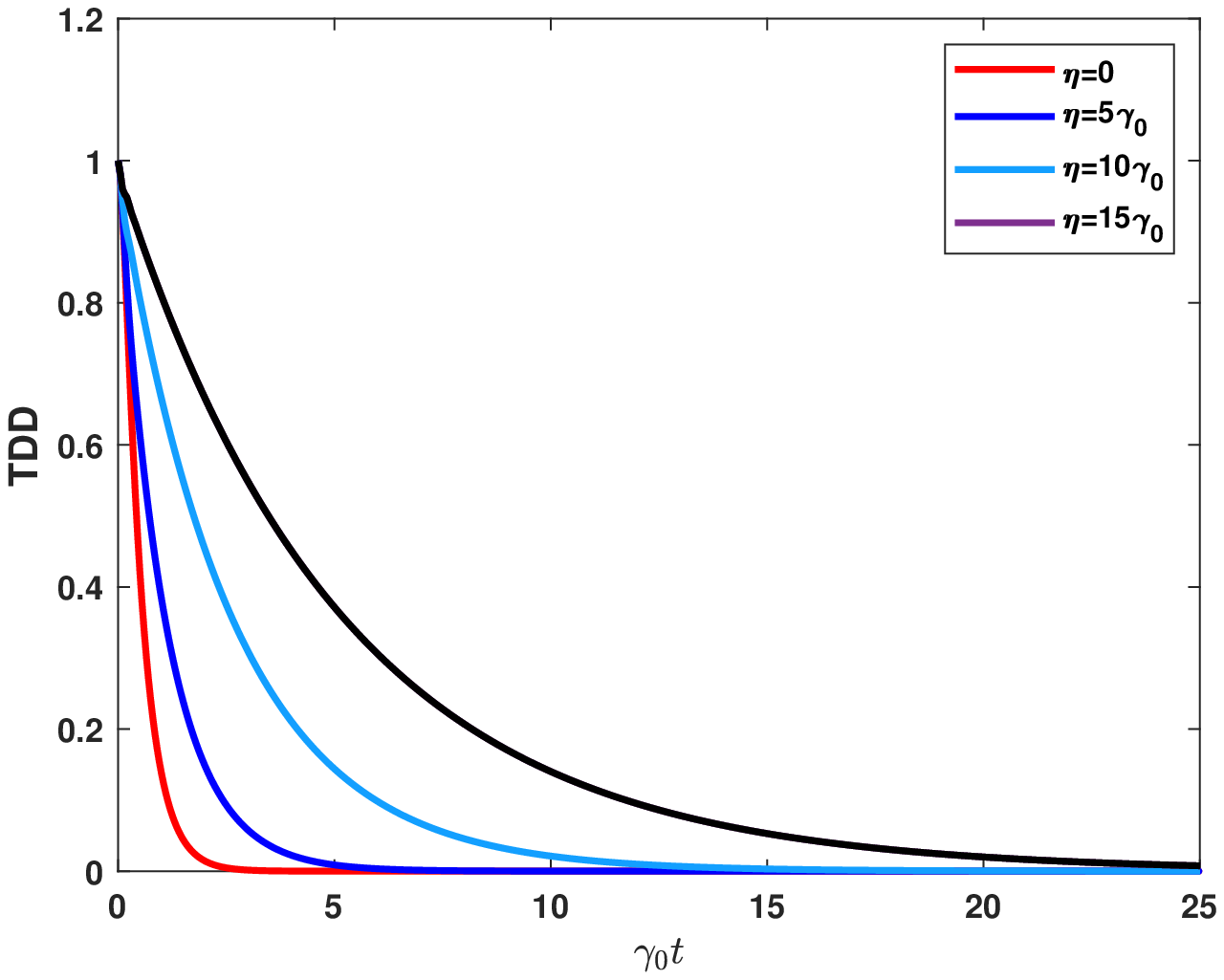}\label{fig3a}}
\subfloat[]{\includegraphics[width=0.5\textwidth]{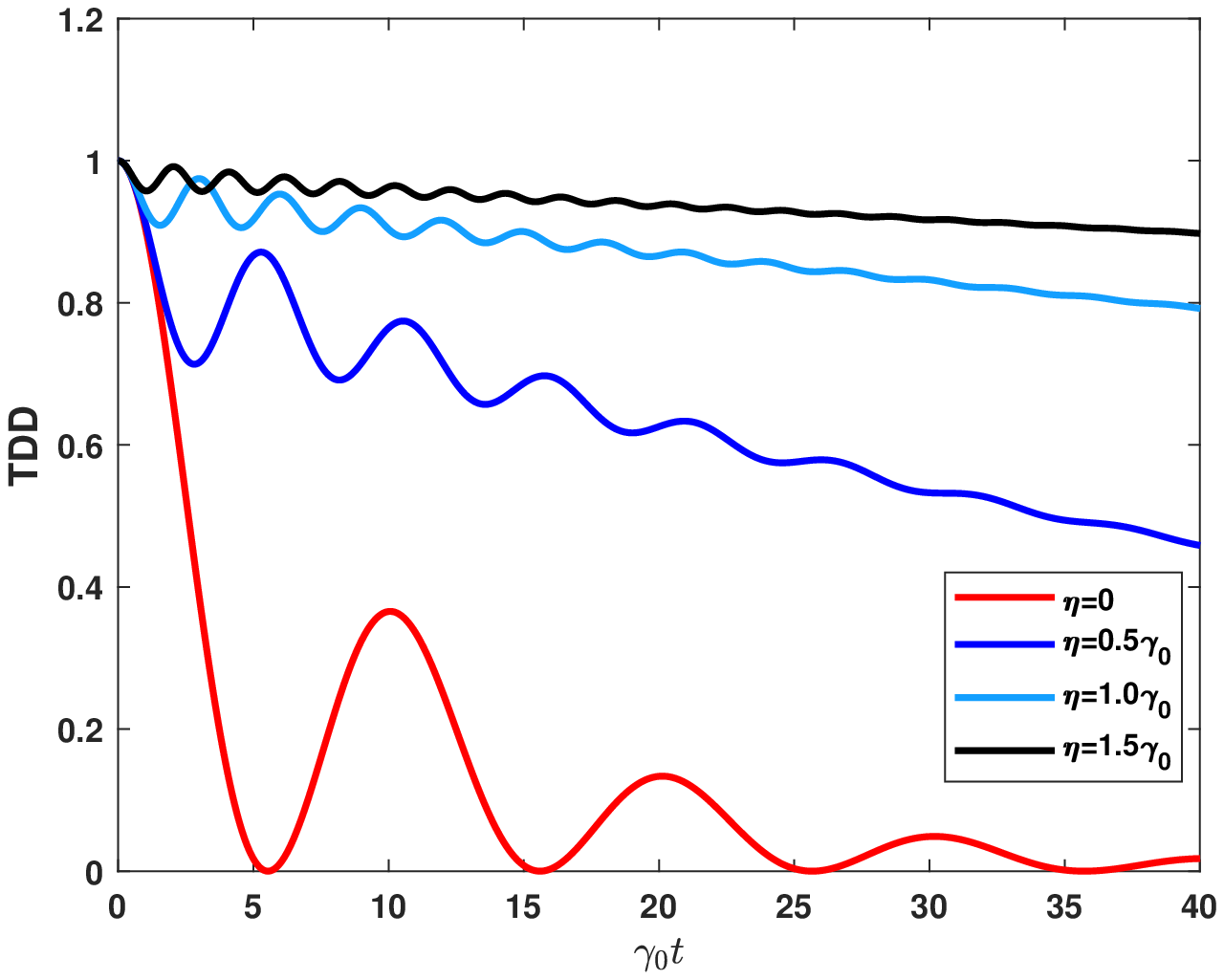}\label{fig3b}}
\caption{The time evolution of TDD with scaled time $\gamma_{0} t$ with state parameter $x=\frac{1}{\sqrt{2}}$ in (a) Markovian reservoir for $\Lambda=10\gamma_0$ with the values of Stark shift ($\eta=0,5\gamma_0,10\gamma_0,15\gamma_0$) and (b) non Markovian reservoir for $\Lambda=0.1\gamma_0$ with $\eta=0, 0.5\gamma_0, 1.0\gamma_0, 1.5\gamma_0$ respectively.}
\label{fig3}
\end{figure}
\begin{figure}[hbt!]
     \centering
\subfloat[]{\includegraphics[width=0.5\textwidth]{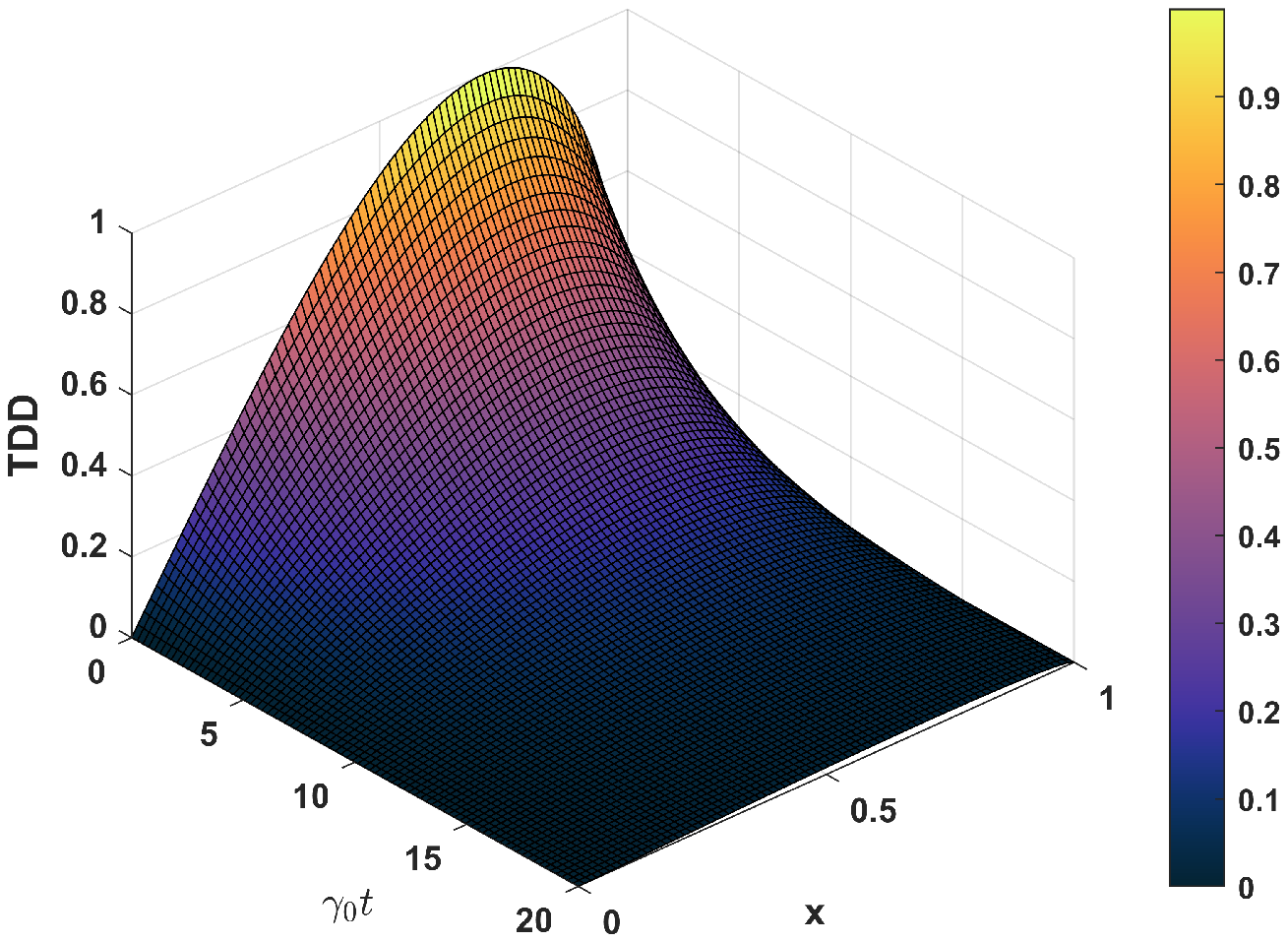}\label{fig4a}}
\subfloat[]{\includegraphics[width=0.5\textwidth]{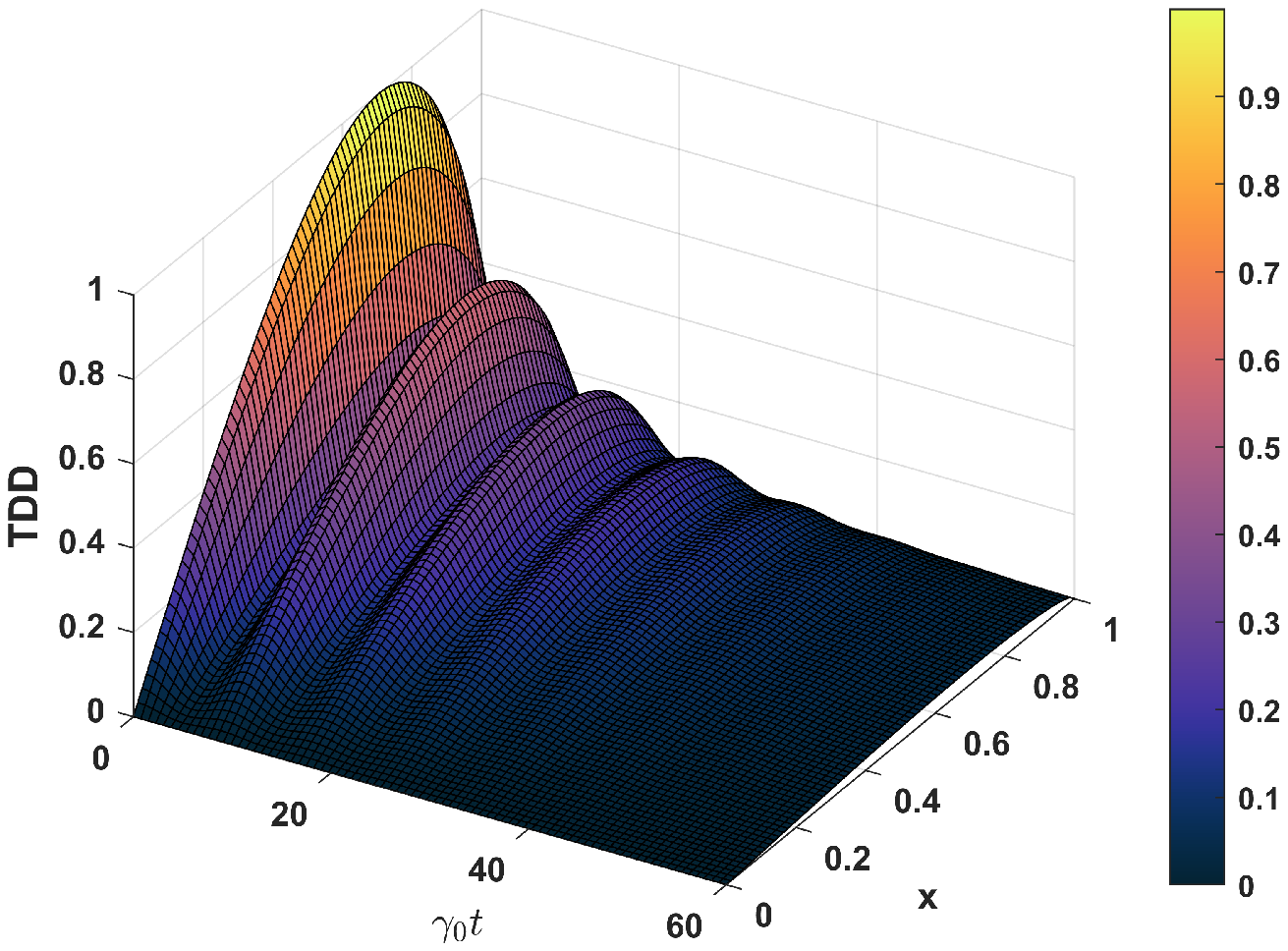}\label{fig4b}}
\caption{The surface plot of TDD with scaled time $\gamma_{0} t$ and state parameter $x$ in (a) in Markovian reservoir for $\Lambda=10\gamma_0$, with the value of Stark shift ($\eta=15\gamma_0$) and (b) in non Markovian reservoir for $\Lambda=0.1\gamma_0$, with the value of Stark shift ($\eta=0.2\gamma_0$).}
\label{fig4}
\end{figure}

The time variation of TDD in the presence of Stark shift is depicted in Fig.\ref{fig3}. 
TDD sustains for a longer period of time with an increase in the Stark shift parameter $(\eta)$.

The surface plots of TDD with scaled time and state parameter are shown in Fig. \ref{fig4}. It is evident from the plot that for maximally entangled input state ($x=\frac{1}{\sqrt{2}})$, TDD has maximum value $1$.
\subsection{\textbf{Dynamics of Local Quantum Uncertainty}}

The calculation of LQU requires one to find the eigenvalues of the matrix $M$:
$$
[M]_{i j} \equiv \operatorname{tr}\left\{\sqrt{\rho}\left(\sigma_{i} \otimes I_{B}\right) \sqrt{\rho}\left(\sigma_{j} \otimes I_{B}\right)\right\},
$$
with elements,
\begin{equation}
    M_{11} = M_{22} =  \frac{2|b_1(t)|^{2}(1 - |b_1(t)|^2 - |b_2(t)|^2)^{1/2}}{(|b_1(t)|^2 + |b_2(t)|^2)^{1/2}},
\end{equation}
\begin{equation}
    M_{33} = \frac{(|b_1(t)|^{2} + |b_2(t)|^{2} - 4|b_1(t)|^{2}|b_2(t)|^{2})}{(|b_1(t)|^{2} + |b_2(t)|^{2})}.
\end{equation}

The elements of the diagonal matrix are the eigenvalues themselves, the expression of LQU is given by
\begin{equation}
\mathcal{Q}(\rho)=1-\max(M_{11}, M_{22},M_{33}).
\end{equation}
\begin{figure}[hbt!]
     \centering
\subfloat[]{\includegraphics[width=0.5\textwidth]{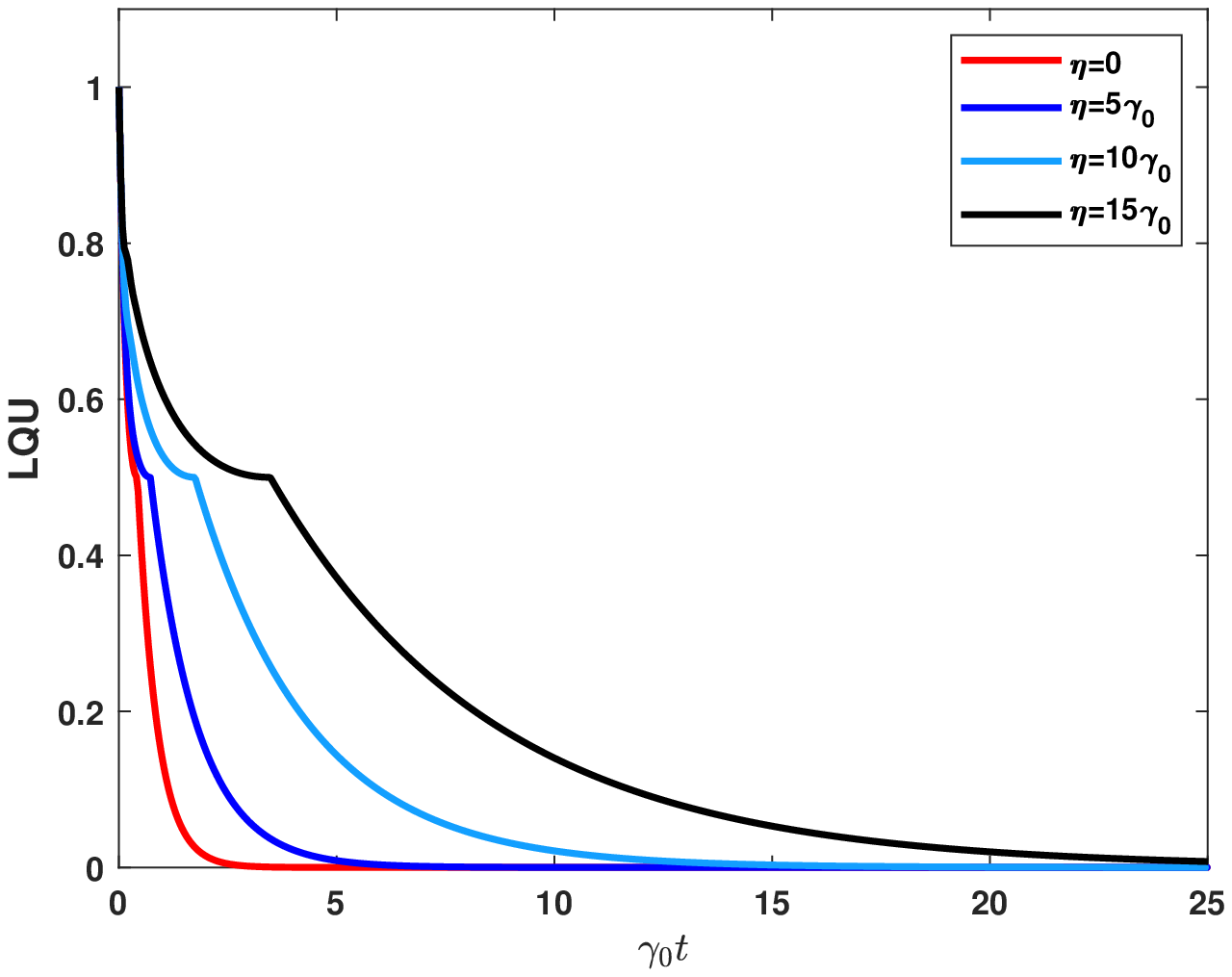}\label{fig5a}}
\subfloat[]{\includegraphics[width=0.5\textwidth]{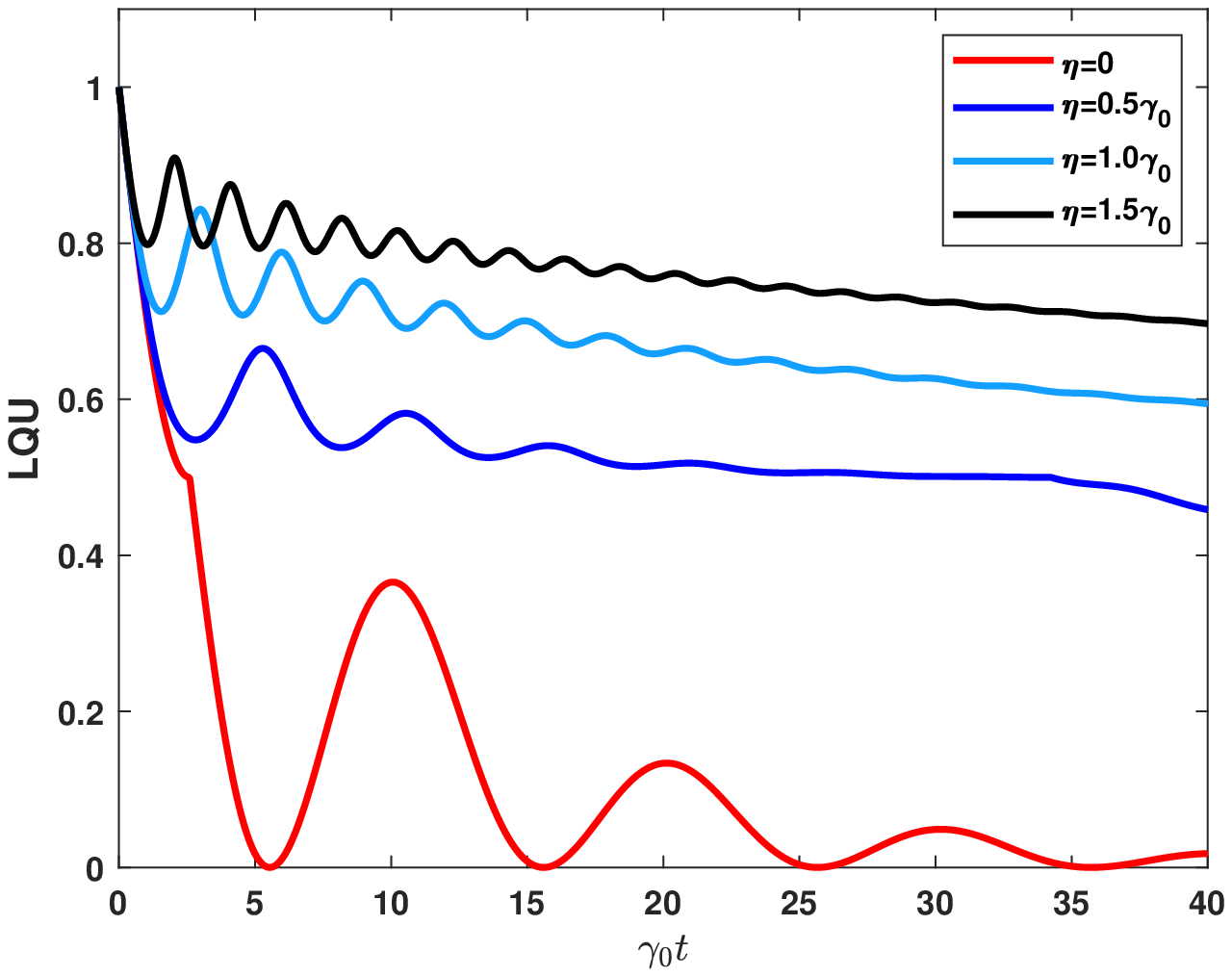}\label{fig5b}}
\caption{The time evolution of LQU with scaled time $\gamma_{0} t$ with $x=\frac{1}{\sqrt{2}}$ in (a) Markovian reservoir for $\Lambda=10\gamma_0$ with the values of Stark shift ($\eta=0,5\gamma_0,10\gamma_0,15\gamma_0$) and (b) non-Markovian reservoir for $\Lambda=0.1\gamma_0$, with $\eta=0, 0.5\gamma_0, 1.0\gamma_0, 1.5\gamma_0$, respectively.}
\label{fig5}
\end{figure}
\begin{figure}[hbt!]
     \centering
\subfloat[]{\includegraphics[width=0.5\textwidth]{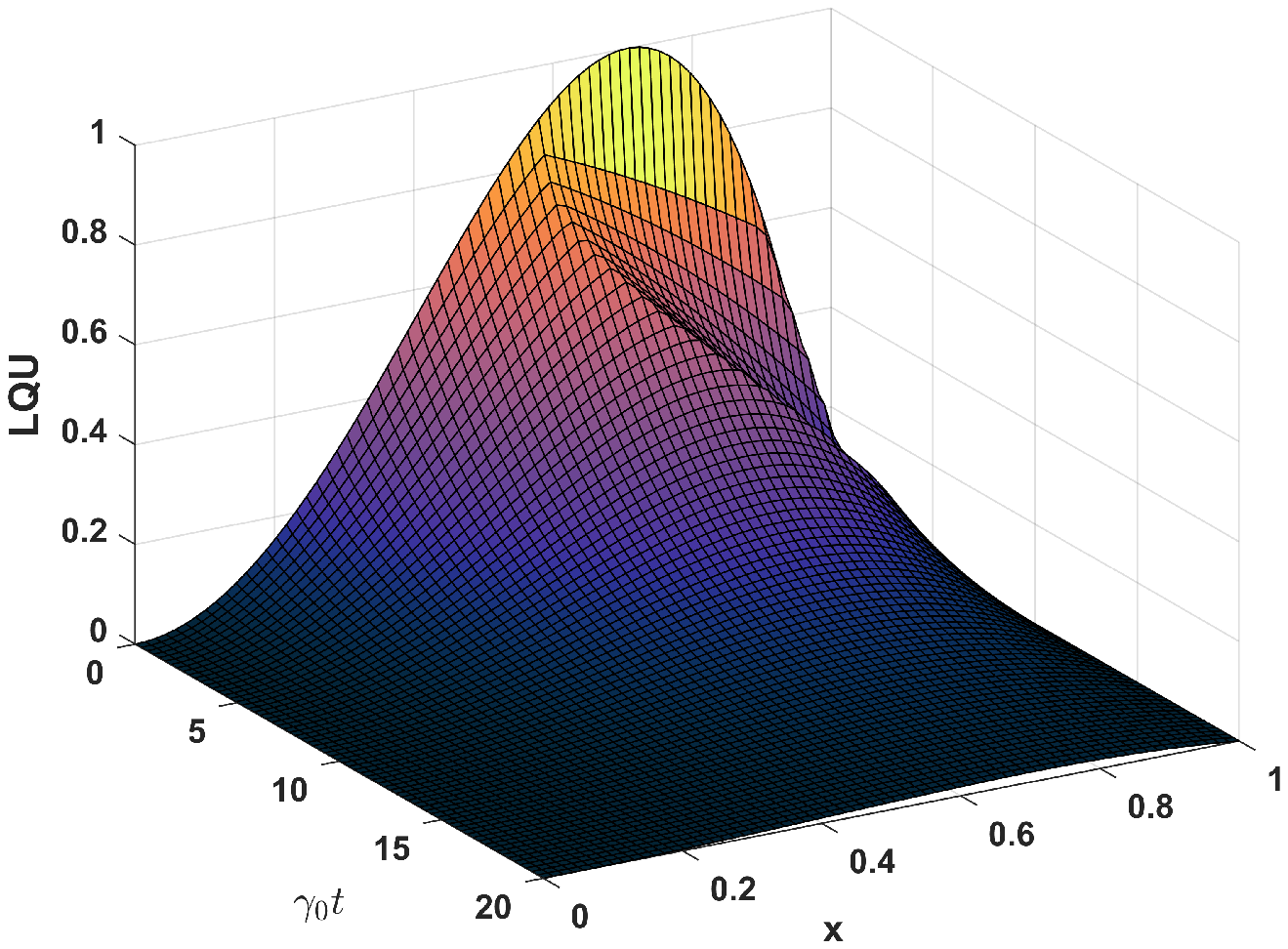}\label{fig6a}}
\subfloat[]{\includegraphics[width=0.5\textwidth]{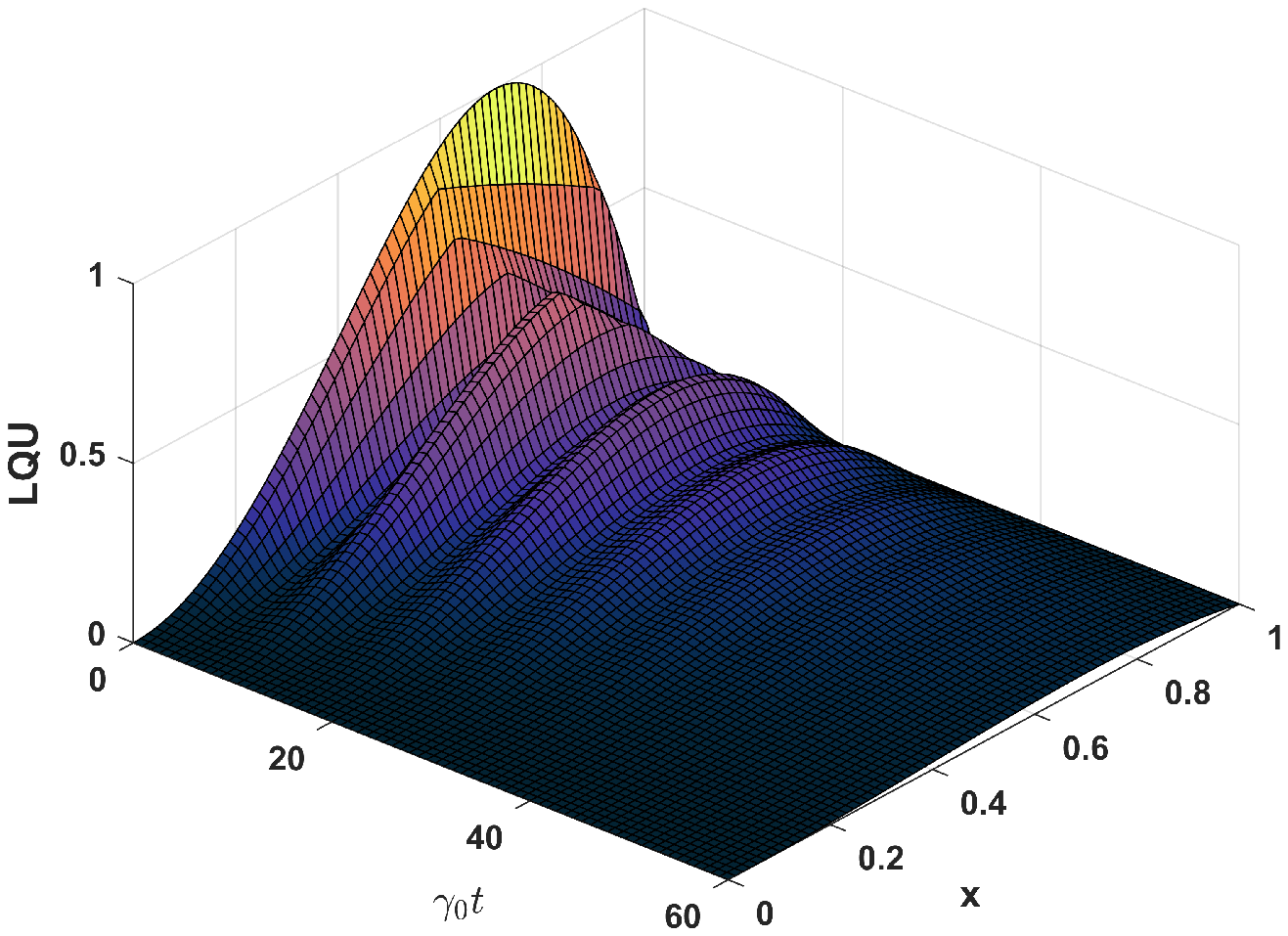}\label{fig6b}}
\caption{The surface plot of LQU with scaled time $\gamma_{0} t$ and state parameter $x$ in (a) in Markovian reservoir for $\Lambda=10\gamma_0$ with the value of Stark shift ($\eta=15\gamma_0$) and (b) in non Markovian reservoir for $\Lambda=0.1\gamma_0$ with the value of Stark shift ($\eta=0.2\gamma_0$).}
\label{fig6}
\end{figure}

In Fig. \ref{fig5}, we have shown the time variation of local quantum uncertainty in the presence of Stark shift. Similar to previous cases, in the Markovian region, we see the exponential decay of TDD. However, we can observe from the dynamics of LQU that it is accompanied with a sudden change in its behavior, which is in contrast to the dynamics of BDE and TDD. This behavior is a result of the maximization process involved in the step of evaluating LQU and happens at a time when $M_{11}=M_{33}$. It must be noted that the sudden change phenomenon of quantum correlations is associated with various physical phenomena like classical and quantum phase transitions \cite{Céleri2017}.

In the non-Markovian region, we observe a damped oscillatory behavior. For each value of $\Lambda$, LQU is enhanced with the increment of Stark shift parameter $\eta$. In the case non-Markovian regime, the significant effect of $\eta$ is observed similar to BDE and TDD. The surface plot of LQU with scaled time and state parameter is shown in Fig.\ref{fig6}. From the surface plot, we observe a region of sudden change as result of the maximization process.
\section{Dynamics of quantum correlation when the mode of environment has one photon initially}
In the previous Section \ref{Sec3}, we have considered the scenario in which the environment of the atoms is in vacuum state, suggesting the absence of photons in the initial state of the environment. We showed that this assumption simplifies to the case where the dynamics of the quantum correlations show dependence on only one of the Stark shift parameters, i,e. '$\eta$' and with the increase of this parameter, we observe an enhancement of quantum correlations. In this section, we deal with the case when both the environment corresponding to the atoms are in the excited state having one photon each.
 Now, we suppose that the initial state of the system environment is $|\Phi^{\prime}(0)\rangle$, where $j^{th}$ mode of each atom's environment has one photon:  
\begin{equation}
|\Phi^{\prime}(0)\rangle=\left(x|0\rangle_{P}|1\rangle_{Q}+\sqrt{1-x^{2}}|1\rangle_{P}|0\rangle_{Q}\right) \otimes\left|1_{j_{1}}\right\rangle_{\Omega_{1}} \mid 1_{j_{2}}\rangle_{\Omega_{2}},
\end{equation}
where $\ket{1_{j_m}}_{\Omega_m}$ indicates that there is one photon in $j^{th}$ mode of the $m^{th}$ environment. 
The joint state of the system and environment at time t can be written as
\begin{equation}
 \begin{split}\label{PHit}   
|\Phi'(t)\rangle=(b'_1(t)\ket{1}_P\ket{0}_Q+b'_2(t)\ket{0}_P\ket{1}_Q)\otimes\ket{1_{j_1}}_{\Omega_1}\ket{1_{j2}}_{\Omega_2}+\sum_{j_{1}} b'_{j_{1}}(t)|0\rangle_{P}|0\rangle_{Q}\left|3_{j_{1}}\right\rangle_{\Omega_{1}}\left|1_{j_{2}}\right\rangle_{\Omega_{2}}\\
+\sum_{j_{2}} b'_{j}(t)|0\rangle_{P}|O\rangle_{Q}\left|1_{j_{1}}\right\rangle_{\Omega_{1}}\left|3_{j_2}\right\rangle_{\Omega_{2}},
\end{split}
\end{equation}
where $\ket{3_{j_m}}_{\Omega_m}$ indicates that there are three photons in $j^{\text{th}}$ mode of the $m^{\text{th}}$ environment. In the interaction picture, the effective Hamiltonian of the system is given by Eq. \ref{Hint}. The time dependent Schrödinger equation in the interaction picture,
$\hat{H}_{int} \ket{\Phi'(t)} = \mathrm{i}\frac{\partial}{\partial t}\ket{\Phi'(t)}$ gives the following differential equations:
\begin{equation}\label{Eq32}
\dot{b'}_{j}(t)=-\mathrm{i} \sqrt{6} \sum_{j_{m}} \gamma_{j_{m}} b'_{j_{m}}(t) \mathrm{e}^{\mathrm{i}\left(\omega_{0}-2\omega_{j_m}\right)t}-3\mathrm{i}\eta {b'}_{j}(t)-\mathrm{i} \xi {b'}_{j}(t)
\end{equation},
and 
\begin{equation}\label{Eq33}
\dot{b'}_{j_m}(t)=-\mathrm{i} \sqrt{6} \gamma_{j_{m}}^{*} b'_{j}(t) \mathrm{e}^{-\mathrm{i}\left(\omega_{0}-2\omega_{j_{m}}\right) t}-4\mathrm{i}  \eta b'_{j_{m}}(t), m=1,2.
\end{equation}
Where, for simplicity we have assumed that, $\eta_{j_1}=\eta_{j_2}=\eta$ ,  $\xi_{j_1}=\xi_{j_2}=\xi$ and $\omega_{j_1}=\omega_{j_2}=\omega_j$. In the process of solving the Schrödinger equation, we note that $\hat{a}_{j_{1}}^{\dagger} \hat{a}_{j_{1}}\xi_{j_{1}} \hat{S}_{+}^{P} \hat{S}_{-}^{P}|\Phi'(t)\rangle$ and $\hat{a}_{j_{2}}^{\dagger} \hat{a}_{j_{2}}\xi_{j_{2}} \hat{S}_{+}^{Q} \hat{S}_{-}^{Q}|\Phi'(t)\rangle$ are nonzero terms. So, the dynamics of quantum correlations also depend on the stark shift parameter $\xi_1$ and $\xi_2$, when the initial mode of the environment is not at vacuum state. We got the following differential equation of probability amplitude using the same approach as before:
\begin{equation}\label{Eq34}
\dot{b'}_{j}(t)=-3\gamma_0\Lambda \int_{0}^{t}\exp{[(\Lambda+4\mathrm{i}\eta)(t-t^\prime)]}  b'_{j}\left(t^{\prime}\right) \mathrm{~d} t^{\prime}-\mathrm{i}(\xi+3\eta){b'}_{j}(t).
\end{equation}
In deriving the above differential equation, the discrete sum over the reservoir modes is replaced by the integral form, $\sum_{k}\left|\gamma_{j}\right|^{2} \rightarrow \int \mathrm{d} \omega_{j} J\left(\omega_{j}\right)$, where $J\left(\omega_{j}\right)$ represents the  Lorentzian spectral density of electric field given in Eq. \ref{Eq9}.
We can find the exact solution $b'_1(t)$ and $b'_2(t)$ by using Laplace technique on Eq. \ref{Eq34}. We obtain the following solutions:
\begin{equation}
b'_1(t)= x \Theta(t), \quad b'_2(t)=\sqrt{1-x^{2}} \Theta(t),
\end{equation}
where,
\begin{equation}
\Theta(t)=e^{-\frac{(\Lambda+ \mathrm{i}(\xi+7\eta) t}{2}}\left[\cosh \left(\frac{\sigma t}{2}\right)+\frac{\Lambda+\mathrm{i}(\eta-\xi)}{\sigma} \sinh \left(\frac{\sigma t}{2}\right)\right],
\end{equation}
with
\begin{equation}
\sigma=\sqrt{-12 \gamma_{0} \Lambda-4\mathrm{i}(\Lambda+4\mathrm{i}\eta)(\xi+3\eta)+(\Lambda+ \mathrm{i} (\xi+7\eta))^{2}}.
\end{equation}

We can write the above state in terms of scaled time '$\gamma_{0}t$', if we write, $\Lambda=l_{1}\gamma_{0}$, $\xi=l_{2}\gamma_{0}$ and $\eta=l_{3}\gamma_{0}$ where $l_{1},l_{2}$ and $l_{3}$ are some numbers.
The state $\Theta(t)$ in terms of scaled time is given by
\begin{equation}
\Theta(\gamma_{0}t)=e^{-\frac{(l_{1}+ \mathrm{i}(l_{2}+7l_{3}) \gamma_{0}t}{2}}\left[\cosh \left(\frac{\sigma_{1} t}{2}\right)+\frac{l_{1}+\mathrm{i}(l_{3}-l_{2})}{\sigma_{1}} \sinh \left(\frac{\sigma_{1} \gamma_{0}t}{2}\right)\right],
\end{equation}
where,
\begin{equation}
\sigma_{1}=\sqrt{-12 l_{1}-4\mathrm{i}(l_{1}+4\mathrm{i}l_{3})(l_{2}+3l_{3})+(l_{1}+ \mathrm{i} (l_{2}+7l_{3}))^{2}}.
\end{equation}

The probability amplitude $b^\prime_1(t)$ and $b'_2(t)$ exhibit different behaviours based on how well a system is coupled to its surroundings.

\begin{figure}[hbt!]\label{last}
     \centering
\subfloat[]{\includegraphics[width=0.33\textwidth]{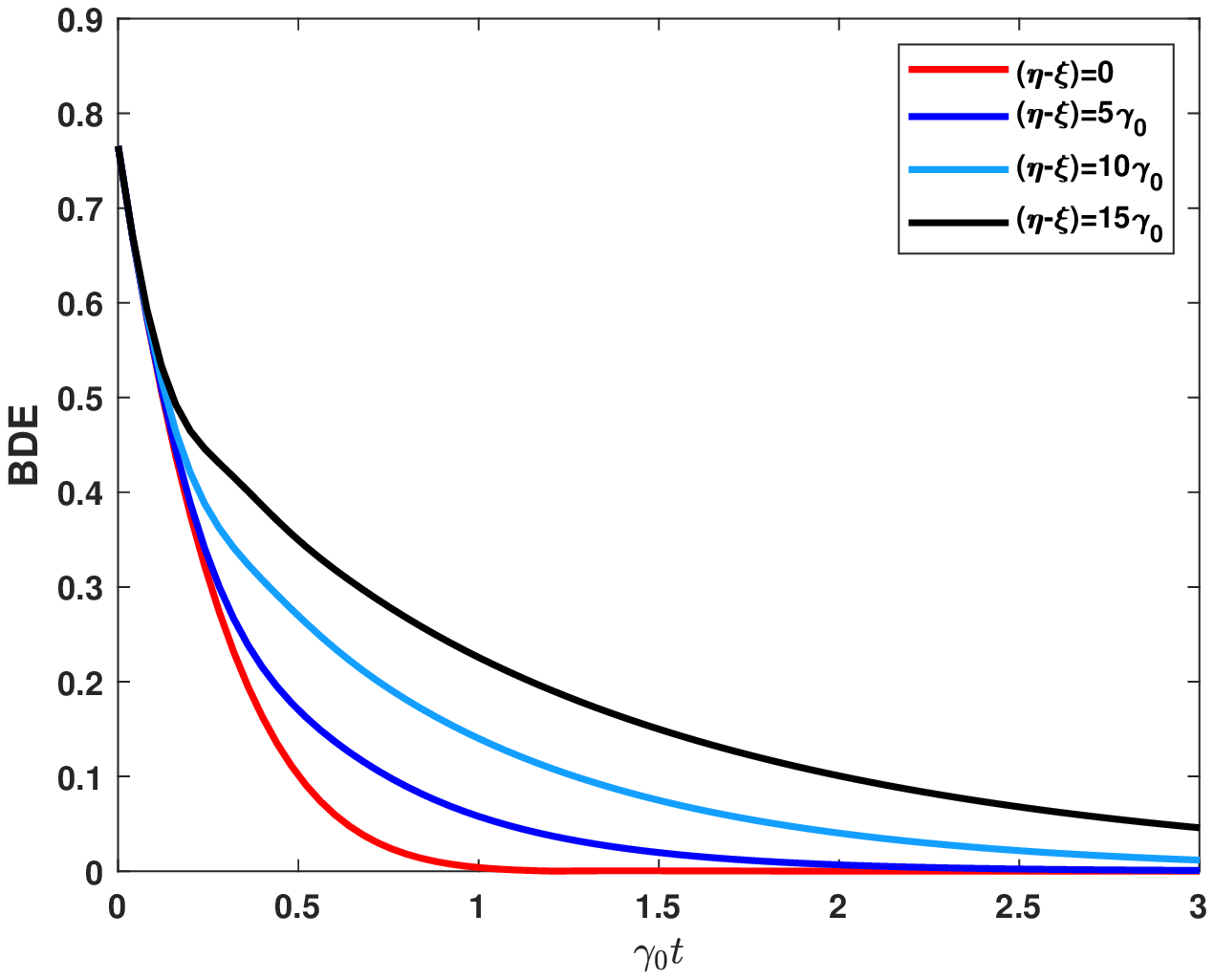}\label{fig7a}}
\subfloat[]{\includegraphics[width=0.33\textwidth]{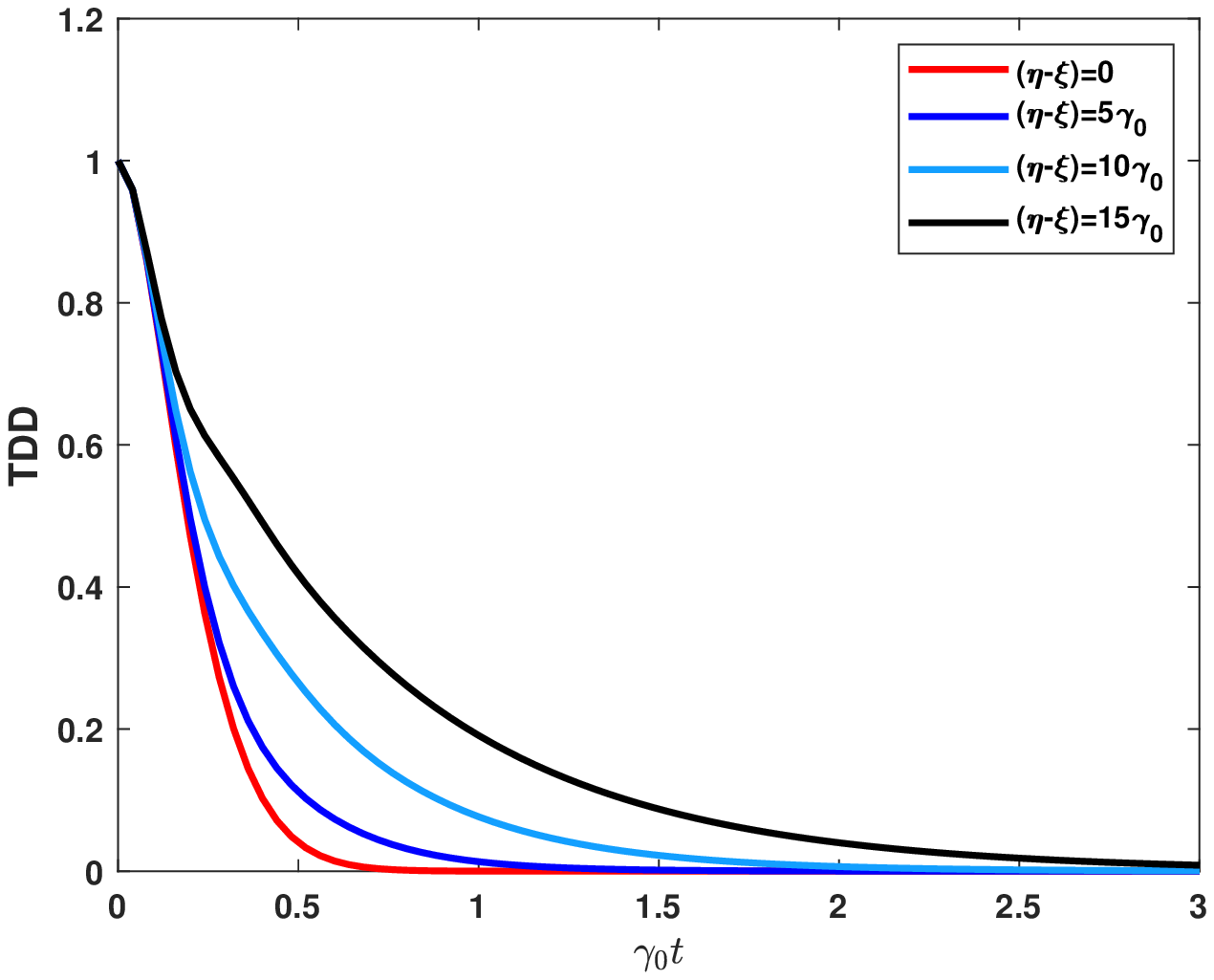}\label{fig7b}}
\subfloat[]{\includegraphics[width=0.33\textwidth]{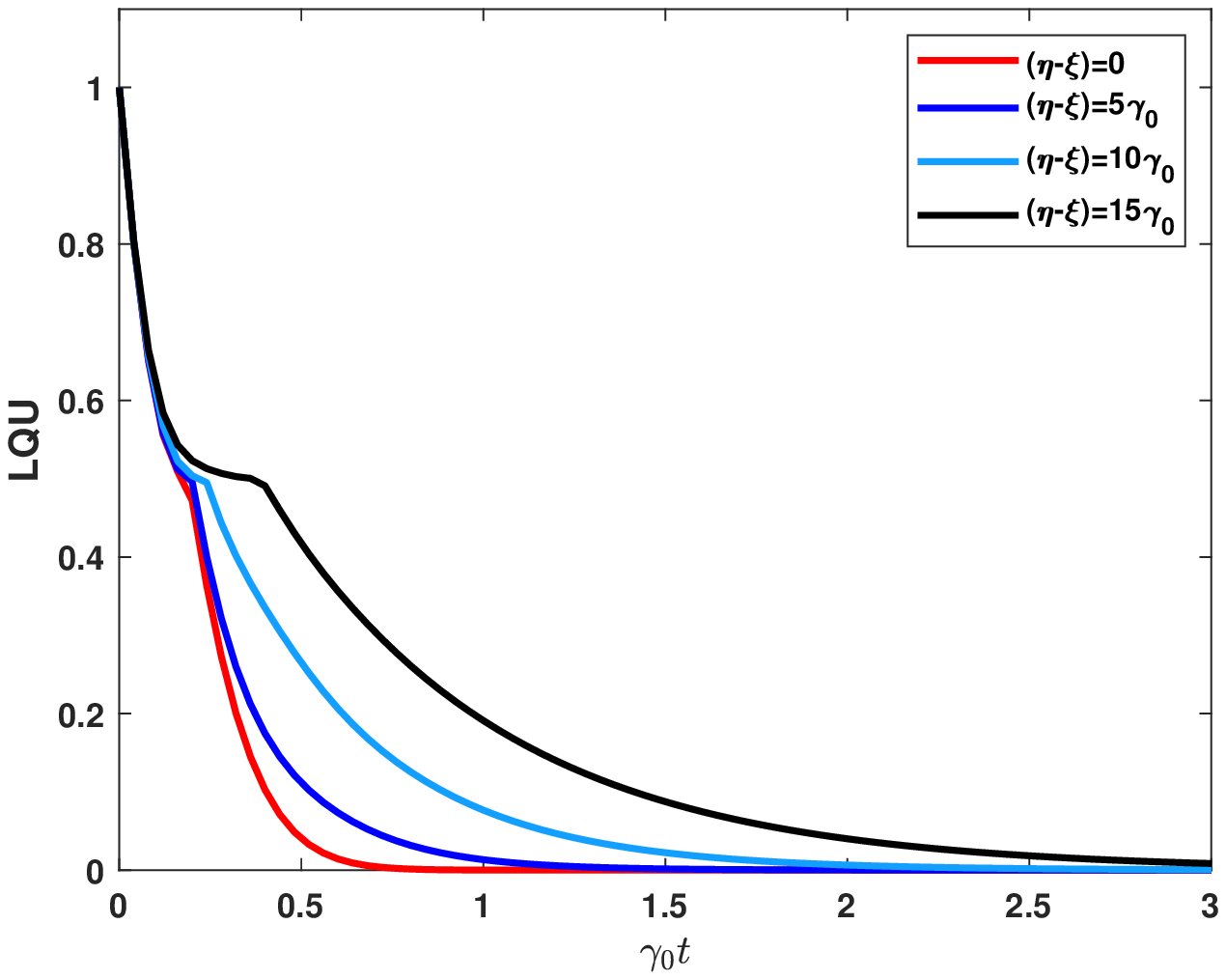}\label{fig7c}}
\newline
\subfloat[]{\includegraphics[width=0.33\textwidth]{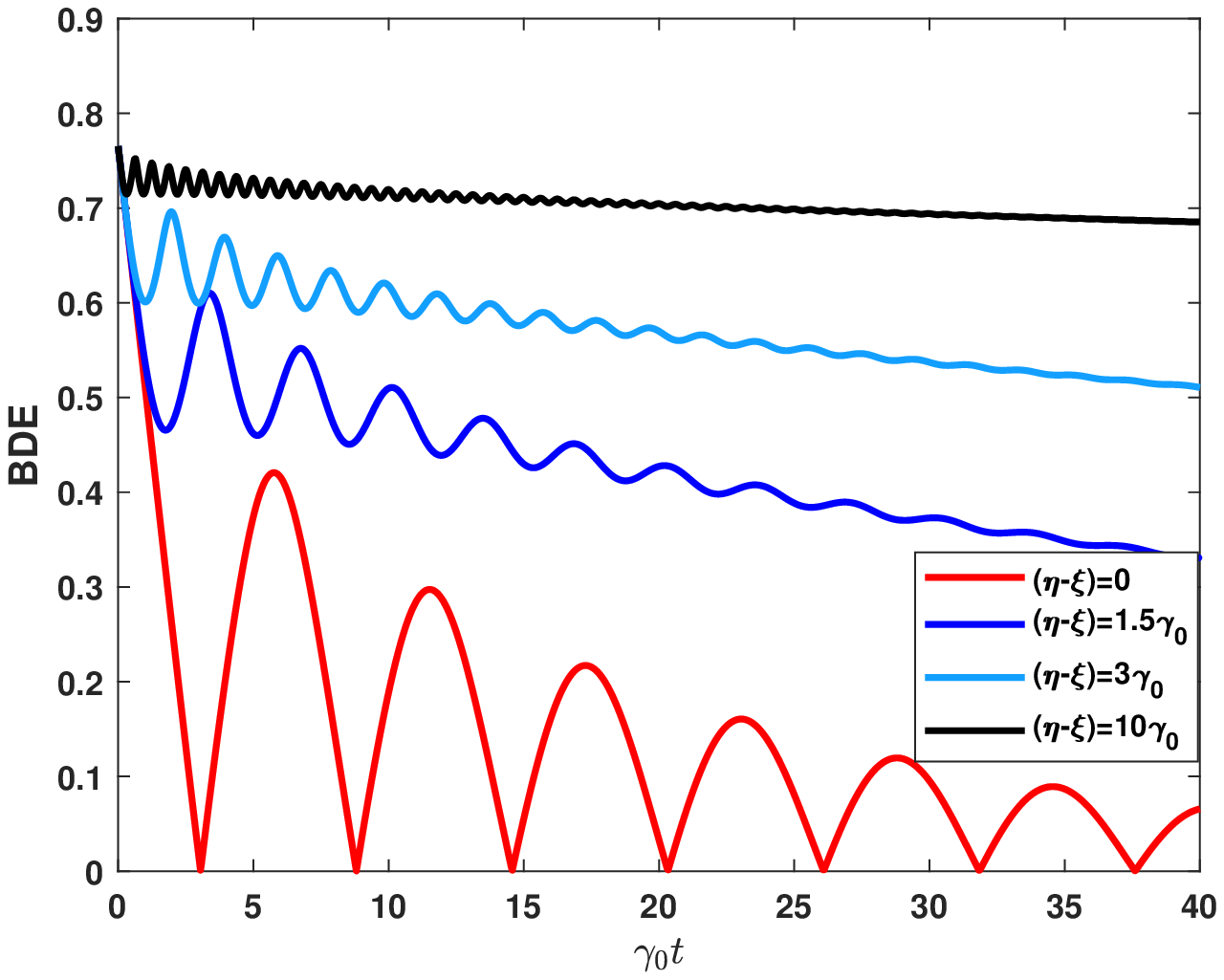}\label{fig7d}}
\subfloat[]{\includegraphics[width=0.33\textwidth]{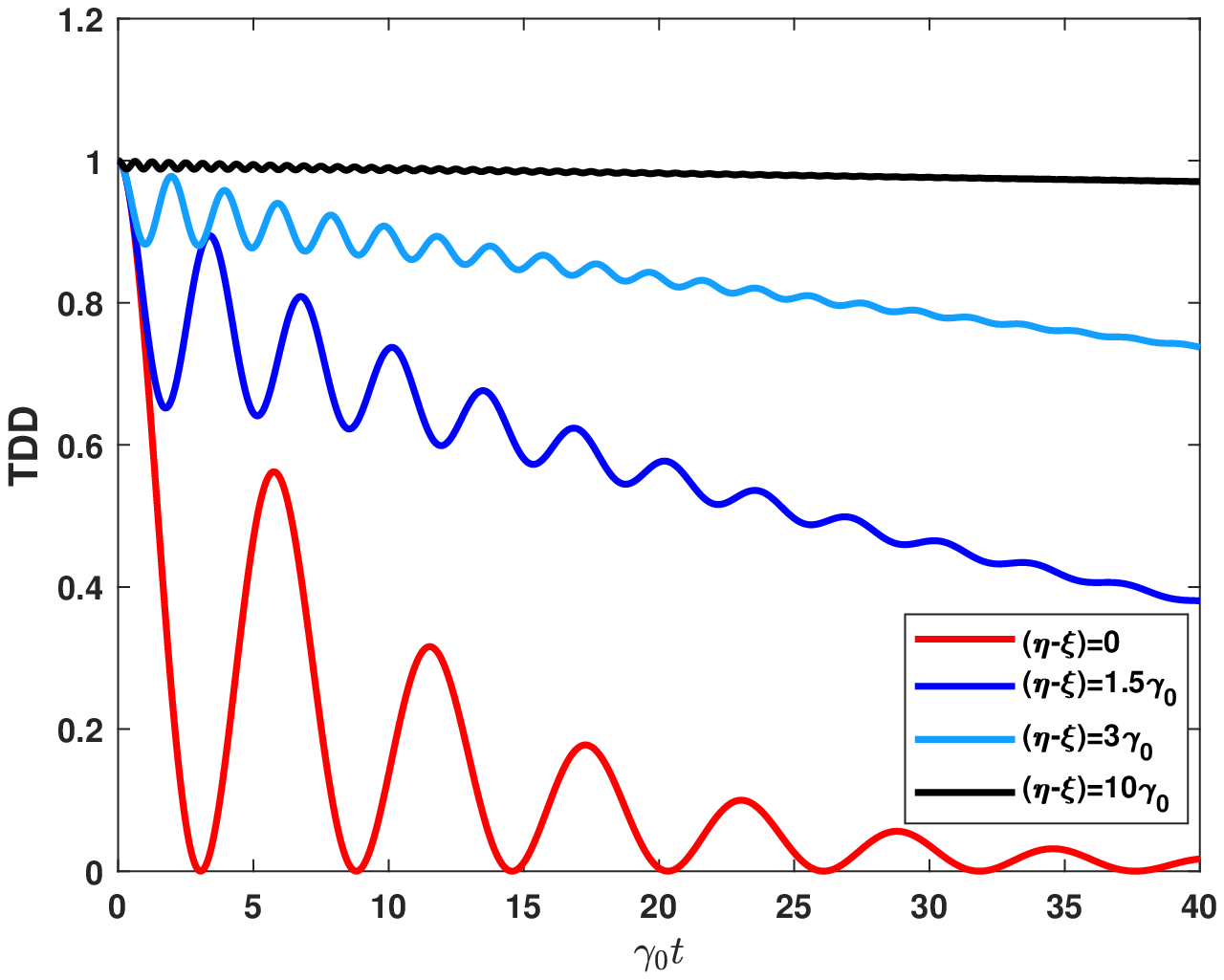}\label{fig7e}}
\subfloat[]{\includegraphics[width=0.33\textwidth]{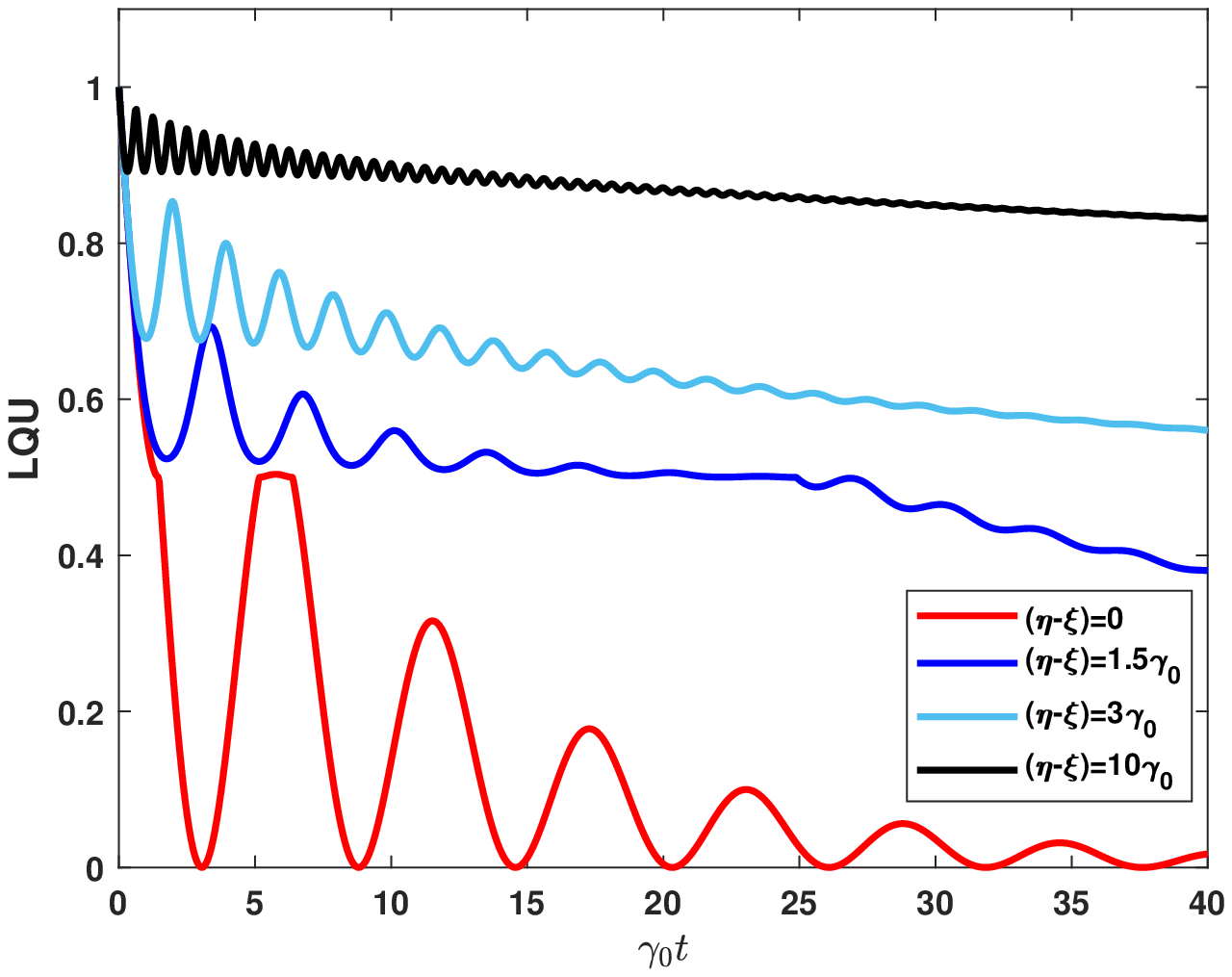}\label{fig7f}}
\caption{The time evolution of BDE, TDD and LQU with scaled time $\gamma_{0} t$ with state parameter $x=\frac{1}{\sqrt{2}}$. Subfigures \ref{fig7a}, \ref{fig7b} and \ref{fig7c} represent  Markovian dynamics for $\Lambda=10\gamma_0$ with the difference of Stark shift parameter $(\eta-\xi)=0,5\gamma_0,10\gamma_0,15\gamma_0$ and subfigures \ref{fig7d}, \ref{fig7e} and \ref{fig7f} represent non Markovian dynamics for $\Lambda=0.1\gamma_0$ with $(\eta-\xi)=0, 1.5\gamma_0, 3.0\gamma_0, 10\gamma_0$ respectively.}
\label{fig7}
\end{figure}

Now, in the presence of the Stark shift effect, we examine the Markovian and non-Markovian dynamics of Bures distance entanglement, trace distance discord, and local quantum uncertainty, which are shown in Figs. \ref{fig7a}, \ref{fig7b}, \ref{fig7c} and Figs. \ref{fig7d}, \ref{fig7e},
\ref{fig7f}.
We have numerically computed the BDE, TDD, and LQU for an initially maximally entangled state and plotted them against scaled time $\gamma_{0} t$ in Fig. \ref{last} for various Stark shift parameters, similar to the vacuum environment field taken into consideration in section \ref{Sec3}.
We found the dynamics of the quantum correlations depend on both the stark shift parameter $\eta$ and $\xi$. The term $(\Lambda+\mathrm{i}(\eta-\xi))/\sigma$ in $\Theta(t)$ plays a significant role in determining the positive effect of Stark shift on the dynamics of quantum correlation measures. From the plots below, we can see that by increasing the value of the difference of the Stark shift parameters $(\eta-\xi)$, the quantum correlations present in the system can be protected effectively and on decreasing the value of $(\eta-\xi)$, we observe quick decay of BDE, TDD and LQU.
The quantum correlations between two atoms in this case decay similarly as in the previous scenario when the environments were in their vacuum state, however, the correlations diminish rapidly when the environments are in the excited state initially. It is important to note that when spectral width ($\Lambda$) decreases, indicating strong non-Markovianity, quantum correlations persist for a longer period of time.
For this initial condition also, we  observe sudden change behaviour in LQU, which is absent in BDE and TDD. 
\section{Conclusion}
In conclusion, we have investigated the dynamics of quantum correlations based on Bures norm, Schatten-1 norm, and local quantum uncertainty for two two-level atoms coupled to dissipative reservoirs at zero temperature in the presence of the Stark effect. It is shown that the quantum correlations can be adequately protected by tuning the magnitude of the Stark shift in the dissipative reservoirs. The quantum correlations in Markovian reservoirs dissipate quickly in comparison to non-Markovian reservoirs. In non-Markovian reservoirs, information flows from the environment back to the system. As a result, the correlations steadily deteriorate. In the presence of the Stark effect, quantum correlations is considerably enhanced by tweaking non-Markovianity or lowering the spectral width. The quantum correlation measures decay with small oscillations for low values of the Stark shift. For high values of the Stark shift, the influence of dissipation is reduced by the Stark effect in non-Markovian reservoirs. It is worth mentioning that quantum correlations quantified by local quantum uncertainty are accompanied by a sudden change phenomenon which is not the case with other correlations based on Bures norm and trace norm. We have looked at the dynamics under two distinct initial conditions of the environment. We have assumed that the initial state of the environment is in the ground state in the first case and the first excited state in the second case. Compared to the first initial condition, which shows the role of only one of the Stark shift parameters, the second initial condition demonstrates the role of both the Stark shift parameters. Our findings have implications for sustaining and controlling quantum correlations beyond entanglement in experiments and the possible use of more general correlation quantifiers in quantum information processing tasks.

\section{Acknowledgements}
 NKC and PKP acknowledge the financial support from DST, India through Grant No. DST/ICPS/QuST/Theme1/2019/2020-21/01. RS acknowledges institute fellowship provided by IISER Kolkata.
\printbibliography

@article{guo2015examining,
  title={"Examining quantum correlations in the XY spin chain by local quantum uncertainty"},
  author={Guo, Jin-Liang and Wei, Jin-Long and Qin, Wan and Mu, Qing-Xia},
  journal={Quantum Information Processing},
  volume={14},
  number={4},
  year={2015},
  publisher={Springer}
}

@article{almeida2007environment,
  title={Environment-induced sudden death of entanglement},
  author={Almeida, Marcelo P and de Melo, Fernando and Hor-Meyll, Malena and Salles, Alejo and Walborn, SP and Ribeiro, PH Souto and Davidovich, Luiz},
  journal={Science},
  volume={316},
  number={5824},
  year={2007},
  publisher={American Association for the Advancement of Science}
}

@article{yu2009sudden,
  title={Sudden death of entanglement},
  author={Yu, Ting and Eberly, JH},
  journal={Science},
  volume={323},
  number={5914},
  year={2009},
  publisher={American Association for the Advancement of Science}
}

@InBook{Joos2009,
author={Joos, Erich and Greenberger, Daniel
and Hentschel, Klaus
and Weinert, Friedel},
title={Compendium of Quantum Physics},
year="2009",
publisher="Springer Berlin Heidelberg",
address="Berlin, Heidelberg",
isbn="978-3-540-70626-7",
doi="10.1007/978-3-540-70626-7_180",
url="https://doi.org/10.1007/978-3-540-70626-7_180"
}

@article{kim2012protecting,
  title={Protecting entanglement from decoherence using weak measurement and quantum measurement reversal},
  author={Kim, Yong-Su and Lee, Jong-Chan and Kwon, Osung and Kim, Yoon-Ho},
  journal={Nature Physics},
  volume={8},
  number={2},
  year={2012},
  publisher={Nature Publishing Group}
}

@article{sun2010reversing,
  title={Reversing entanglement change by a weak measurement},
  author={Sun, Qingqing and Al-Amri, M and Davidovich, Luiz and Zubairy, M Suhail},
  journal={Physical Review A},
  volume={82},
  number={5},
  year={2010},
  publisher={APS}
}

@InBook{Lidar2003,
author="Lidar, Daniel A.
and Birgitta Whaley, K.",
editor="Benatti, Fabio
and Floreanini, Roberto",
title={Decoherence-Free Subspaces and Subsystems},
bookTitle="Irreversible Quantum Dynamics",
year="2003",
publisher="Springer Berlin Heidelberg",
address="Berlin, Heidelberg",
isbn="978-3-540-44874-7",
doi="10.1007/3-540-44874-8_5",
url="https://doi.org/10.1007/3-540-44874-8_5"
}

@article{flores2015two,
  title={Two qubit entanglement preservation through the addition of qubits},
  author={Flores, MM and Galapon, EA},
  journal={Annals of Physics},
  volume={354},
  year={2015},
  publisher={Elsevier}
}

@article{mortezapour2017protecting,
  title={Protecting entanglement by adjusting the velocities of moving qubits inside non-Markovian environments},
  author={Mortezapour, Ali and Borji, Mahdi Ahmadi and Franco, Rosario Lo},
  journal={Laser Physics Letters},
  volume={14},
  number={5},
  year={2017},
  publisher={IOP Publishing}
}

@article{horodecki2009quantum,
  title={Quantum entanglement},
  author={Horodecki, Ryszard and Horodecki, Pawe{\l} and Horodecki, Micha{\l} and Horodecki, Karol},
  journal={Reviews of Modern Physics},
  volume={81},
  number={2},
  year={2009},
  publisher={APS}
}

@article{bennett1993teleporting,
  title={Teleporting an unknown quantum state via dual classical and Einstein-Podolsky-Rosen channels},
  author={Bennett, Charles H and Brassard, Gilles and Cr{\'e}peau, Claude and Jozsa, Richard and Peres, Asher and Wootters, William K},
  journal={Physical Review Letters},
  volume={70},
  number={13},
  year={1993},
  publisher={APS}
}

@article{ekert1991quantum,
  title={Quantum cryptography based on Bell’s theorem},
  author={Ekert, Artur K},
  journal={Physical Review Letters},
  volume={67},
  number={6},
  year={1991},
  publisher={APS}
}

@article{maccone2013intuitive,
  title={Intuitive reason for the usefulness of entanglement in quantum metrology},
  author={Maccone, Lorenzo},
  journal={Physical Review A},
  volume={88},
  number={4},
  year={2013},
  publisher={APS}
}

@article{bell1964einstein,
  title={On the Einstein Podolsky Rosen paradox},
  author={Bell, John S},
  journal={Physics Physique Fizika},
  volume={1},
  number={3},
  year={1964},
  publisher={APS}
}

@article{bera2017quantum,
  title={Quantum discord and its allies: a review of recent progress},
  author={Bera, Anindita and Das, Tamoghna and Sadhukhan, Debasis and Roy, Sudipto Singha and De, Aditi Sen and Sen, Ujjwal},
  journal={Reports on Progress in Physics},
  volume={81},
  number={2},
  year={2017},
  publisher={IOP Publishing}
}

@article{knill1998power,
  title={Power of one bit of quantum information},
  author={Knill, Emanuel and Laflamme, Raymond},
  journal={Physical Review Letters},
  volume={81},
  number={25},
  year={1998},
  publisher={APS}
}

@article{datta2008quantum,
  title={Quantum discord and the power of one qubit},
  author={Datta, Animesh and Shaji, Anil and Caves, Carlton M},
  journal={Physical Review Letters},
  volume={100},
  number={5},
  year={2008},
  publisher={APS}
}

@book{fanchini2017lectures,
  title={Lectures on General Quantum Correlations and Their Applications},
  author={Fanchini, Felipe Fernandes and Pinto, Diogo de Oliveira Soares and Adesso, Gerardo},
  year={2017},
  publisher={Springer}
}

@article{spehner2014quantum,
  title={Quantum correlations and distinguishability of quantum states},
  author={Spehner, Dominique},
  journal={Journal of Mathematical Physics},
  volume={55},
  number={7},
  year={2014},
  publisher={American Institute of Physics}
}

@article{ciccarello2014toward,
  title={Toward computability of trace distance discord},
  author={Ciccarello, F and Tufarelli, T and Giovannetti, V},
  journal={New Journal of Physics},
  volume={16},
  number={1},
  year={2014},
  publisher={IOP Publishing}
}

@article{girolami2013characterizing,
  title={Characterizing nonclassical correlations via local quantum uncertainty},
  author={Girolami, Davide and Tufarelli, Tommaso and Adesso, Gerardo},
  journal={Physical Review Letters},
  volume={110},
  number={24},
  year={2013},
  publisher={APS}
}

@article{luo2003wigner,
  title={Wigner-Yanase skew information and uncertainty relations},
  author={Luo, Shunlong},
  journal={Physical Review Letters},
  volume={91},
  number={18},
  year={2003},
  publisher={APS}
}

@article{piani2012problem,
  title={Problem with geometric discord},
  author={Piani, Marco},
  journal={Physical Review A},
  volume={86},
  number={3},
  year={2012},
  publisher={APS}
}

@article{luo2004wigner,
  title={Wigner-Yanase skew information vs. quantum Fisher information},
  author={Luo, Shunlong},
  journal={Proceedings of the American Mathematical Society},
  volume={132},
  number={3},
  year={2004}
}

@article{bellomo2007non,
  title={Non-Markovian effects on the dynamics of entanglement},
  author={Bellomo, Bruno and Franco, R Lo and Compagno, Giuseppe},
  journal={Physical Review Letters},
  volume={99},
  number={16},
  year={2007},
  publisher={APS}
}

@article{ghosh2008control,
  title={Control of atomic entanglement by the dynamic Stark effect},
  author={Ghosh, Biplab and Majumdar, AS and Nayak, N},
  journal={Journal of Physics B: Atomic, Molecular and Optical Physics},
  volume={41},
  number={6},
  year={2008},
  publisher={IOP Publishing}
}

@article{baghshahi2014entanglement,
  title={Entanglement analysis of a two-atom nonlinear Jaynes--Cummings model with nondegenerate two-photon transition, Kerr nonlinearity, and two-mode stark shift},
  author={Baghshahi, HR and Tavassoly, MK and Faghihi, MJ},
  journal={Laser Physics},
  volume={24},
  number={12},
  year={2014},
  publisher={IOP Publishing}
}

@article{golkar2018dynamics,
  title={Dynamics and maintenance of bipartite entanglement via the stark shift effect inside dissipative reservoirs},
  author={Golkar, S and Tavassoly, MK},
  journal={Laser Physics Letters},
  volume={15},
  number={3},
  year={2018},
  publisher={IOP Publishing}
}

@article{wootters1998entanglement,
  title={Entanglement of formation of an arbitrary state of two qubits},
  author={Wootters, William K},
  journal={Physical Review Letters},
  volume={80},
  number={10},
  year={1998},
  publisher={APS}
}

@article{streltsov2010linking,
  title={Linking a distance measure of entanglement to its convex roof},
  author={Streltsov, Alexander and Kampermann, Hermann and Bru{\ss}, Dagmar},
  journal={New Journal of Physics},
  volume={12},
  number={12},
  year={2010},
  publisher={IOP Publishing}
}

@article{fano1957description,
  title={Description of states in quantum mechanics by density matrix and operator techniques},
  author={Fano, Ugo},
  journal={Reviews of modern physics},
  volume={29},
  number={1},
  year={1957},
  publisher={APS}
}

@article{luo2019quantifying,
  title={Quantifying nonclassicality via Wigner-Yanase skew information},
  author={Luo, Shunlong and Zhang, Yue},
  journal={Physical Review A},
  volume={100},
  number={3},
  year={2019},
  publisher={APS}
}

@article{puri1988quantum,
  title={Quantum electrodynamics of an atom making two-photon transitions in an ideal cavity},
  author={Puri, RR and Bullough, RK},
  journal={JOSA B},
  volume={5},
  number={10},
  year={1988},
  publisher={Optical Society of America}
}

@InBook{breuer2002theory,
  title={The theory of open quantum systems},
  author={Breuer, Heinz-Peter and Petruccione, Francesco and others},
  year={2002},
  publisher={Oxford University Press on Demand}
}

@article{spohn1980kinetic,
  title={Kinetic equations from Hamiltonian dynamics: Markovian limits},
  author={Spohn, Herbert},
  journal={Reviews of Modern Physics},
  volume={52},
  number={3},
  year={1980},
  publisher={APS}
}

@article{chen2021dynamics,
  title={Dynamics of local quantum uncertainty and local quantum Fisher information for a two-qubit system driven by classical phase noisy laser},
  author={Chen, Lu-ping and Guo, You-neng},
  journal={Journal of Modern Optics},
  volume={68},
  number={4},
  year={2021},
  publisher={Taylor \& Francis}
}

@article{khedif2018local,
  title={Local quantum uncertainty and trace distance discord dynamics for two-qubit X states embedded in non-Markovian environment},
  author={Khedif, Youssef and Daoud, Mohammed},
  journal={International Journal of Modern Physics B},
  volume={32},
  number={20},
  year={2018},
  publisher={World Scientific}
}

@article{slaoui2018dynamics,
  title={The dynamics of local quantum uncertainty and trace distance discord for two-qubit X states under decoherence: a comparative study},
  author={Slaoui, Abdallah and Daoud, Mohammed and Laamara, R Ahl},
  journal={Quantum Information Processing},
  volume={17},
  number={7},
  year={2018},
  publisher={Springer}
}

@article{dalton2001theory,
  title={Theory of pseudomodes in quantum optical processes},
  author={Dalton, BJ and Barnett, Stephen M and Garraway, BM},
  journal={Physical Review A},
  volume={64},
  number={5},
  year={2001},
  publisher={APS}
}

@article{mohamed2021quantum,
  title={Quantum Fisher Information and Bures Distance Correlations of Coupled Two Charge-Qubits Inside a Coherent Cavity with the Intrinsic Decoherence},
  author={Mohamed, Abdel-Baset A and Khalil, Eied and Selim, Mahmoud M and Eleuch, Hichem and others},
  journal={Symmetry},
  volume={13},
  number={2},
  year={2021},
  publisher={Multidisciplinary Digital Publishing Institute}
}

@InBook{wigner1997information,
  author={Wigner, Eugene P and Yanase, Mutsuo M},
  title={Part I: Particles and Fields. Part II: Foundations of Quantum Mechanics},
  year={1997},
  publisher={Springer}
}

@InBook{Céleri2017,
author={C{\'e}leri, Lucas C.
and Maziero, Jonas},
title={The Sudden Change Phenomenon of Quantum Discord},
bookTitle={Lectures on General Quantum Correlations and their Applications},
year={2017},
publisher={Springer International Publishing},
address={Cham}
}

@article{adesso2016measures,
  title={Measures and applications of quantum correlations},
  author={Adesso, Gerardo and Bromley, Thomas R and Cianciaruso, Marco},
  journal={Journal of Physics A: Mathematical and Theoretical},
  volume={49},
  number={47},
  year={2016},
  publisher={IOP Publishing}
}

@article{haas2006two,
  title={Two-photon excitation dynamics in bound two-body Coulomb systems including ac Stark shift and ionization},
  author={Haas, Martin and Jentschura, Ulrich D and Keitel, Christoph H and Kolachevsky, Nikolai and Herrmann, Maximilian and Fendel, Peter and Fischer, M and Udem, Th and Holzwarth, Ronald and H{\"a}nsch, TW and others},
  journal={Physical Review A},
  volume={73},
  number={5},
  year={2006},
  publisher={APS}
}

@article{agarwal2004dc,
  title={dc-field-induced enhancement and inhibition of spontaneous emission in a cavity},
  author={Agarwal, GS and Pathak, PK},
  journal={Physical Review A},
  volume={70},
  number={2},
  year={2004},
  publisher={APS}
}

@article{singh2015experimental,
  title={Experimental estimation of discord in an antiferromagnetic Heisenberg compound ${Cu(NO_{3})_{2}.2.5H_{2}O}$},
  author={Singh, Harkirat and Chakraborty, Tanmoy and Panigrahi, Prasanta K and Mitra, Chiranjib},
  journal={Quantum Information Processing},
  volume={14},
  number={3},
  year={2015},
  publisher={Springer}
}

@article{yu2004finite,
  title={Finite-time disentanglement via spontaneous emission},
  author={Yu, Ting and Eberly, JH},
  journal={Physical Review Letters},
  volume={93},
  number={14},
  year={2004},
  publisher={APS}
}

@article{eberly2007end,
author = {J. H. Eberly  and Ting Yu },
title = {The End of an Entanglement},
journal = {Science},
volume = {316},
number = {5824},
year = {2007},
}

@article{chandra2022dissipative,
  title={Dissipative dynamics of quantum correlation quantifiers under decoherence channels},
  author={Chandra, Nitish Kumar and Bhosale, Sarang S and Panigrahi, Prasanta K},
  journal={The European Physical Journal Plus},
  volume={137},
  number={4},
  year={2022},
  publisher={Springer}
}

@article{sk2022protecting,
  title={Protecting quantum coherence and entanglement in a correlated environment},
  author={Sk, Rajiuddin and Panigrahi, Prasanta K},
  journal={Physica A: Statistical Mechanics and its Applications},
  volume={596},
  year={2022},
  publisher={Elsevier}
}
\end{document}